\documentclass[12pt]{article}
\usepackage{latexsym}
\usepackage{amssymb}
\usepackage{graphicx}
\usepackage{xypic}
\usepackage{epsf}

\def\K{{\cal{K}}}

\def\reals{{\Re}}

\def\K{{\cal{K}}}

\newcommand{\TL}{Temperley--Lieb}

\newcommand{\Sph}{{\mathbb S}}
\newcommand{\Dom}{\mbox{dom }}

\providecommand{\noglossaryignore}[1]{}
\newcommand{\globalglossaryentry}[3]{\makebox[1.5in][l]{\tt $\backslash${#1}} 
\makebox[1.1in][l]{{$#2$}} \makebox[2.5in][l]{{#3}}\newline} 
\newcommand{\newcommandabbreviation}[3]{\newcommand{#1}{#2}%
\noglossaryignore{\globalglossaryentry{#3}{#2}{}}}
\newcommand{\renewcommandabbreviation}[3]{\renewcommand{#1}{#2}%
\noglossaryignore{\globalglossaryentry{#3}{#2}{}}}
\newcommand{\newcommandmacro}[4]{\newcommand{#1}{#2}%
\noglossaryignore{\globalglossaryentry{#3}{#2}{#4}}}
\newcommand{\gge}[3]{\noglossaryignore{\globalglossaryentry{#1}{#2}{#3}}}
\newcommand{\myaddress}%
{\parbox{3in}{\footnotesize \begin{center} 
Mathematics Department, City University, \\  
Northampton Square, London EC1V 0HB, UK.\end{center}}}
   
\newcounter{minidef}[section]

\newcounter{minicapt}

\newtheorem{theo}{Theorem}         
\newtheorem{de}{Definition}     \newtheorem{pr}{Proposition}

\newtheorem{claim}{Claim}
\noglossaryignore{GREEK ETC.\newline}
\newcommandabbreviation{\e}{\epsilon}{e}        
\newcommandabbreviation{\lam}{\lambda}{lam}  
\newcommandabbreviation{\la}{\langle}{la}        
\newcommandabbreviation{\ran}{\rangle}{ran}
\newcommandabbreviation{\ha}{\#}{ha}             
\newcommandabbreviation{\rmap}{\rightarrow}{rmap}
\newcommandabbreviation{\aaa}{\alpha}{aaa}        
\newcommandabbreviation{\ab}{\alpha,\beta}{ab}
\newcommandabbreviation{\aab}{a(\ab )}{aab}       
\noglossaryignore{\newline RINGS\newline}
\newcommandabbreviation{\HH}{H \!\!\! I}{HH}              
\newcommandabbreviation{\C}{\mathbb C}{C}
\newcommandabbreviation{\N}{\mathbb N}{N}  
\newcommandabbreviation{\Z}{\mathbb Z}{Z}     
\renewcommandabbreviation{\Re}{\mathbb R}{Re}
\newcommandabbreviation{\R}{{\mathbb R}}{R}
\newcommandabbreviation{\Q}{\mathbb Q }{Q}
\renewcommandabbreviation{\H}{\mathbb H }{H}
\noglossaryignore{\newline SYMMETRIC GROUP\newline}
\def\Sym(#1){\Sigma(#1)}                  
\gge{Sym(-)}{\Sym(-)}{symmetric group on - objects}
\def\Sy(#1){\Sigma_{#1}}                  
\gge{Sy(-)}{\Sy(-)}{symmetric group irreducible -}
\def\sym(#1){\mbox{\LARGE s}(#1)}       
\gge{sym(-)}{\sym(-)}{symmetric group on - objects (variant)}
\def\sy(#1){\mbox{\LARGE s}({#1})}       
\gge{sy(-)}{\sy(-)}{symmetric group irreducible - (variant)}
\newcommandmacro{\cs}{\C \, \sy(n)}{cs}{symmetric group algebra over $\C$}
\noglossaryignore{\newline PARTITIONS/SETS\newline}
\newcommand{\Nset}[1]{\underline{#1}}
\gge{Nset\{-\}}{\Nset{-}}{set of natural numbers to -}
\def\nset(#1){ \{ #1 \}_{ \underline{n} }}
\gge{nset(-)}{\nset(-)}{a set $-\times\Nset$}
\def\ul(#1){_{\underline{#1}}}            
\gge{ul(-)}{{}\ul(-)}{subscript underline -}
\def\Ee(#1){{\bf E}_{#1}}                 
\gge{Ee(-)}{\Ee(-)}{set of equivalence relations on set -}
\def\Eee(#1){{\bf E}_{\{ #1 \}_{\underline{n}}}}  
\gge{Eee(-)}{\Eee(-)}{ditto for nset}
\def\Een(#1,#2){{\bf E}_{\{ #1 \}_{\underline{#2}}}}  
\def\Ssn(#1,#2){{\bf S}_{\{ #1 \}_{\underline{#2}}}}  
\def\Ss(#1){{\bf S}_{#1}}                 
\def\Sss(#1){{\bf S}_{\{ #1 \}_{\underline{n}}}}  
\def\bbc(#1){((\beta_1)(\beta_2)...(\beta_{#1}))}     
\newcommandmacro{\Ln}{{\Gamma}^{n}}{Ln}{large index set}
\newcommandmacro{\LnQ}{{\Gamma}^{n}_Q}{LnQ}{index set}
\newcommandmacro{\Zz}{\zeta}{Zz}{`shape' function}
        
\noglossaryignore{\newline PARTITION ALGEBRA\newline}
\def\ka(#1){\kappa_{#1}}                  
\def\Sm(#1){\Sigma_{#1}}                  
\newcommandmacro{\com}{\bullet}{com}{bullet composition}
\newcommandmacro{\enm}{\; e^n(\! m\! ) \;}{enm}{product of idempotents}
\def\Ai(#1){ A^{ #1 \cdot } }             
\def\Aij(#1,#2){ A^{ #1  #2 } }           
\newcommandmacro{\One}{\mbox{\bf $1 \!\!\! 1$}}{One}{algebra unit 1}

\newcommandmacro{\Bp}{B_p}{Bp}{partition basis}
\def\Bb(#1){B_p[#1]}                      
\def\Pp(#1){P_n[#1]}                      
\def\Ps(#1){P_n[#1] \! /}                 
\newcommandmacro{\Ph}{\hat{P}}{Ph}{P hat  algebra}
\def\Is(#1){\sim^{#1}}                    
\noglossaryignore{\newline MODULES\newline}
\def\Wm(#1){{\cal S}_{#1}}                
\gge{Wm(-)}{\Wm(-)}{Weyl module with index -}
\def\wm(#1,#2){{}_{#1}{\cal S}_{#2}}      
\gge{wm(-1,-)}{\wm(-1,-)}{Weyl module with index -}
\def\Ind(#1,#2,#3){\mbox{Ind}_{#1}^{#2}#3}
\gge{Ind(-1,-2,-)}{\Ind(-1,-2,-)}{induction}
\def\Res(#1,#2,#3){\mbox{Res}_{#1}^{#2}#3}
\gge{Res(-1,-2,-)}{\Res(-1,-2,-)}{restriction}
\newcommandabbreviation{\weyl}{standard}{weyl}
\newcommandabbreviation{\mod}{\mbox{mod}}{mod}
\newcommandabbreviation{\head}{\mbox{head }}{head}
\newcommandabbreviation{\Weyl}{Weyl}{Weyl}
\def\SS(#1){{\cal S}_{#1}}                
\gge{SS(-)}{\SS(-)}{Specht/Weyl module index -}
\def\LL(#1){{\cal L}_{#1}}                
\gge{LL(-)}{\LL(-)}{simple module index -}
\noglossaryignore{\newline FUNCTORS/MAPS\newline}
\newcommandmacro{\Gg}{{\cal G}}{Gg}{G Functor}
\newcommandmacro{\Fg}{{\cal F}}{Fg}{F Functor}
\newcommandmacro{\ra}{\rightarrow}{ra}{}
\def\ses(#1,#2,#3){0\ra #1 \ra #2 \ra #3 \ra 0}  
\gge{ses(1,2,-)}{\ses(1,2,-)}{\hspace{.5in} short exact sequence}
\def\starr(#1){ \stackrel{ #1 }{\longrightarrow} }
\gge{starr(-)}{\starr(-)}{}
\newcommandmacro{\doublerightarrow}{\; -\!\!\! -\!\!\!\!\!\! \gg \;}
{doublerightarrow}{}
\noglossaryignore{\newline PARTITION ALGEBRA MAPS\newline}
\newcommandmacro{\smap}{s}{smap}{`inclusion' map}
\newcommandmacro{\tmap}{t}{tmap}{$ P_n -> S_n$}
\newcommandmacro{\pmap}{\psi}{pmap}{$ S_n -> P_n $}
\noglossaryignore{\newline MISC.\newline}
\def\Amap(#1){{\cal A}_{#1}}              
\gge{Amap(-)}{\Amap(-)}{}
\def\Rr(#1){R_{#1}}                       
\gge{Rr(-)}{\Rr(-)}{restriction of E}
\def\Cr(#1){C_{#1}}                       
\gge{Cr(-)}{\Cr(-)}{restriction of E to N}
\newcommandmacro{\Tm}{{\cal T}}{Tm}{Transfer Matrix}
\def\On(#1){{\cal I}_{#1}}
\gge{On(-)}{\On(-)}{}
\newcommandmacro{\UU}{\underline{\sqcup}}{UU}{}  
\newcommandmacro{\UUU}{\sqcup}{UUU}{}  
\newcommandmacro{\Vq}{V_Q^{\otimes n}}{Vq}{Potts config. space}
\def\bs(#1,#2){\mbox{{\Large $\ast$}}^{#1}_{#2}} 
\gge{bs(-,-)}{\bs(-,-)}{general plumbing multiplier}
\newcommand{\ignore}[1]{}
\newcommand{\hignore}[1]{}
\gge{ignore\{-\}}{\hignore{-}}{ignore argument!}
\def\choo(#1,#2){ \left( \begin{array}{c} #1 \\ #2 \end{array} \right) }
\gge{choo(-1,-)}{\choo(-1,-)}{choose}
\newcommand{\Qed}{$\Box$}
\gge{Qed}{\mbox{\Qed}}{QED}
\def\staq(#1){\stackrel{#1}{=}}           
\gge{staq(-)}{\staq(-)}{}
\def\stam(#1){\stackrel{#1}{\rightarrow}} 
\gge{stam(-)}{\stam(-)}{}
\def\mat{ \left( \begin{array} }    
\def\tam{ \end{array}  \right) }
\gge{mat/tam}{...}{matrix delimiters}
\newcommand{\beq}{\begin{equation} }
\def\eql(#1){ \begin{equation} \label{#1} 
}
\newcommand{\eq}{\end{equation} }
\def\eqal(#1){\begin{eqnarray} \label{#1} }
\def\eqa{\end{eqnarray} }
\def\lab(#1){\label{#1}
}
\def\prl(#1){ \begin{pr} \label{#1} 
}
\def\del(#1){ \begin{de} \label{#1} 
}
   
\gge{smeq\{-\}}{...}{small equation}
  
\gge{fneq\{-\}}{...}{very small equation}
\noglossaryignore{\newline HECKE/BLOB\newline}
\newcommandmacro{\Hnq}{H_n(q)}{Hnq}{ * freestanding symbol}
\newcommandmacro{\Hn}{H_n}{Hn}{      *-mod etc.}
\newcommandmacro{\A}{{\cal A}}{A}{}
\newcommandmacro{\Cwts}{C}{Cwts}{}
\newcommandmacro{\CA}{{\cal A}}{CA}{}

\newcommandmacro{\calA}{{\cal A}}{calA}{}
\newcommandmacro{\modi}{\mbox{Mod} }{modi}{was mod not modi!}
\newcommandmacro{\Wgen}{{\Bbb S}}{Wgen}{}
\def\ol(#1){\overline{#1}}
\newcommandmacro{\st}{\mbox{St}}{st}{}
  
\def\CMult(#1,#2){(#1:#2)}
\def\CM(#1,#2){( #1 : #2 )}
\def\FMult#1,#2{(#1:#2)}
\def\CF#1,#2{(#1:#2)}

\newcommandmacro{\Top}{\mbox{Top}}{Top}{}
\newcommandmacro{\Soc}{\mbox{Soc}}{Soc}{}
\newcommandmacro{\Head}{\mbox{Head}}{Head}{}
\newcommandmacro{\Filt}{{\cal F}}{Filt}{}
\newcommandmacro{\Mod}{\mbox{mod}}{Mod}{}
\newcommandmacro{\Resi}{\mbox{Res }}{Resi}{was without i!}
\newcommandmacro{\Indi}{\mbox{Ind }}{Indi}{was without i!}
  
\def\RR(#1,#2){R^{#1}_{#2}}  
\def\TT(#1,#2){T^{#1}_{#2}}

\def\St{\st}
\newcommandmacro{\Ann}{\mbox{Ann}}{Ann}{}
\newcommandmacro{\Cen}{\mbox{Cen}}{Cen}{}
\newcommandmacro{\End}{\mbox{End}}{End}{}
\newcommandabbreviation{\semisimple}{semisimple}{semisimple}
\newcommandabbreviation{\Bratteli}{Bratteli}{Bratteli}
\newcommandabbreviation{\JBC}{Jones Basic Construction}{JBC}
\newcommandabbreviation{\pa}{partition algebra}{pa}
\newcommandabbreviation{\TM}{transfer matrix}{TM}
\newcommandabbreviation{\PM}{Potts model}{PM}
\newcommandabbreviation{\QSC}{quantum spin chain}{QSC}
\newcommandabbreviation{\Hamiltonian}{Hamiltonian}{Hamiltonian}
\newcommandabbreviation{\YS}{Young symmetrizer}{YS}

\newcommand{\eeq}{\end{equation}}
\newcommand{\bear}{\begin{eqnarray}}
\newcommand{\eear}{\end{eqnarray}}
\newcommand{\bearn}{\begin{eqnarray*}}
\newcommand{\eearn}{\end{eqnarray*}}

\newcommand{\myfig}[2]{\raisebox{#1 in}{ \includegraphics{#2.eps}}}

\newtheorem{para}{Paragraph}[section]
\newtheorem{definition}[para]{Definition}
\newtheorem{proposition}[para]{Proposition}
\newtheorem{lemma}[para]{Lemma}
\newtheorem{theorem}[para]{Theorem}
\newtheorem{remark}[para]{Remark}
\newtheorem{example}[para]{Example}

\newcommand{\theor}[1]{\begin{theorem} #1 \end{theorem}}
\renewcommand{\de}[1]{\begin{definition} #1 \end{definition}}
\renewcommand{\pr}[1]{\begin{proposition} #1 \end{proposition}}
\newcommand{\lm}[1]{\begin{lemma} #1 \end{lemma}}
\newcommand{\ex}[1]{\begin{example}{\rm #1 }\end{example}}
\newcommand{\re}[1]{\begin{remark}{\rm  #1 }\end{remark}}

\newcommand{\oproof}[2]{{\em Proof:} #2 \Qed}
\newcommand{\proof}[1]{{\em Proof:} #1 \Qed}
\newcommand{\ochat}[1]{}

\newcommand{\DV}{\partial}   
\newcommand{\Pt}{P}          

\newcommand{\SD}{S_D^}        

\newcommand{\Sd}{S_{\DV}^}    
\renewcommand{\St}{S^}       
\newcommand{\Sto}{S_{[\!]}^} 
\newcommand{\Smin}{S_{min}^} 
\newcommand{\SSmin}[1]{{S}^{#1}_{\he}}
\newcommand{\SSmins}[1]{{S}^{#1}_{\she}}

\newcommand{\tz}{t}
\newcommand{\DDD}[2]{D^{#1}_{#2}} 

\newcommand{\CD}{concrete diagram}
\newcommand{\CDD}{concrete $\partial$-diagram}
\newcommand{\CLC}{loop configuration}
\newcommand{\CBC}{boundary configuration}

\newcommand{\CCDs}{isotopy classes of \CD s}

\newcommand{\resp}{respectively}

\newcommand{\Ring}{{\mathcal K}}

\newcommand{\Power}{{\mathcal P}}
\newcommand{\Pa}{{\mathbb P}}
\newcommand{\pcirc}{*} 
\newcommand{\CC}[1]{{\mathcal C}^{#1}} 
 
\newcommand{\conn}{p}  
\newcommand{\Rem}{{\mathsf R}} 
\newcommand{\Uni}{{\sf U}}  
\newcommand{\dotcup}{\stackrel{.}{\cup}} 
 
\newcommand{\FFF}{[F,F]}
\newcommand{\bub}{b} 
\newcommand{\genus}{g}

\newcommand{\he}{{\mathfrak h}}
\newcommand{\she}{{\mathfrak{sh}}}
\newcommand{\ii}{{\mathfrak i}}
\newcommand{\si}{{\mathfrak{si}}}
\newcommand{\rr}{{\mathfrak r}}

\newcommand{\simsh}{\sim_{\she}}  
\newcommand{\simh}{\sim_{\he}}  
\newcommand{\simi}{\sim_{\ii}}  
\newcommand{\simsi}{\sim_{\si}}

\newcommand{\Heterotopy}{Heterotopy}

\newcommand{\het}{heterotopy}

\newcommand{\shet}{strong heterotopy}

\newcommand{\asphere}{s}     

\newcommand{\tree}{{\mathcal T}}
\newcommand{\ptree}{{\mathcal T}^p}  
\newcommand{\Graph}{{\mathcal H}}  
\newcommand{\bracket}{{\mathcal B}}
\newcommand{\gtree}{{\mathcal G}}
\newcommand{\chain}{\mbox{ch}}  
\newcommand{\RTS}{require to show}

\newcommand{\swedge}{\nabla}
\newcommand{\cwedge}{\overline{\swedge}}

\newcommand{\Down}{{\mathbb D}}

\newcommand{\futnote}[1]{\footnote{CAN DROP THIS?: #1 }}
\newcommand{\xfutnote}[1]{}

\title{A Temperley--Lieb category for 2-manifolds}\date{}
\author{Marcos Alvarez and P P Martin \\ \myaddress}
\begin{document}

\maketitle

\begin{center}
\bf Abstract
\end{center}
Guided by consideration of problems in 2 and 3 dimensional lattice
model computation, we are led to 
define a number of new categories, and functors
between these categories and the partition category, culminating in the
introduction of two categories generalising the \TL\ category.
We show how to compute practically in these categories, by giving a
combinatorial realisation of their (topological) construction.

\tableofcontents
\section{Introduction}

This work is motivated, on the one hand, by the need implicit in 
certain topological models of magnetic charge \cite{Sorkin77,DiemerHadley99} 
to understand the construction of space manifolds by gluing bounded manifolds
through a common spherical boundary \cite{Alvarez07}, and on 
the other hand by the idea of generalising the planar 
diagram calculus \cite{Weyl46} of the 
Temperley-Lieb algebra \cite{TemperleyLieb71} to higher dimensions. 

The Temperley-Lieb algebra crops up in a wide variety of mathematical and 
physical contexts, from algebraic Lie theory \cite{Jimbo85,Kassel95}, 
statistical mechanics \cite{Baxter82,Martin91}
(and see Section~\ref{S:dichro}), 
knot theory \cite{Kauffman91}, 
conformal field theory \cite{KooSaleur93} and combinatorics, 
to colouring problems \cite{Saleur90}, 
TQFT, Khovanov homology \cite{KhovanovSeidel02} and 
C*-algebras \cite{GoodmandelaHarpeJones89}. 
In most of those contexts the planar diagram calculus can be seen 
as integral to the algebra's involvement. 
In this calculus the algebra (indeed category) has a 
basis of diagrams drawn on a rectangular interval of the plane. The rectangle 
contains non-intersecting lines which end on its upper and lower edges, 
and two diagrams are equivalent if they differ by an edge-preserving ambient 
isotopy. Two diagrams A and B may be composed into a third diagram AB if the 
number of lines in A ending on its lower edge equals the number of lines 
in B ending on its upper edge.

It is of interest in several of the contexts above to try to generalise this
setup to higher dimensions. In particular, several interesting two-dimensional
models have been solved in statistical mechanics by algebraic methods 
\cite{Baxter82,Martin91}, but almost none in dimension three, which is
the dimension most directly relevant to equilibrium physics
(cf. \cite{BazhanovBaxter94,ReggeZecchina00}).

There are a number of forms this 
diagram calculus 
generalisation might take, and various such 
have been considered  
\cite{Martin91,Martin94}. 
Perhaps the most superficially obvious generalisation, composition of open
$m$-manifolds 
embedded in
$(m+1)$-space\footnote{Actually we will argue 
elsewhere that $\scriptstyle m$-manifolds in $\scriptstyle 2m$-space may be 
more interesting in the context of statistical mechanics, but this is even 
harder, so we start here.}, 
has not previously been fully treated from this point of view 
(although see \cite{Martin91,BaezDolan95,BaezLangford97})
for the following reason. Non-intersecting 
lines intersect the (let's say) upper edge of a 
\TL\ diagram $A$ at distinct points, 
and distinct points on $\R$ have a natural order that is preserved by 
ambient isotopy. Thus for diagrams to be composable the 
line counting condition above is sufficient. 
(This is the same as to say that the object set in the category is
the natural numbers $\N_0$.)
If $m=2$ then lines become surfaces and intersect the northern and southern 
plane in loops. Clearly, for two ``diagrams'' to compose the number of loops 
in the juxtaposed layers must match, but this embedding does not order the 
loops, and the matching loop number is not sufficient for two loop 
configurations to be ambient isotopic and hence for diagrams with these 
configurations to be composable
(cf. cobordism and TQFT \cite{Atiyah90}). 
Another associated complication that occurs for $m=2$ is 
that some composable diagrams can be composed in more than one way. 

More than one 
formal resolution of these ambiguities is possible.
To explain a way to choose a `good' resolution we now recall the framework 
from Physics which provides the motivating example.
An overview of the paper follows in Section~\ref{S:over}.

\subsection{The Potts model/dichromatic polynomial paradigm}\label{S:dichro}
\begin{figure}
\includegraphics{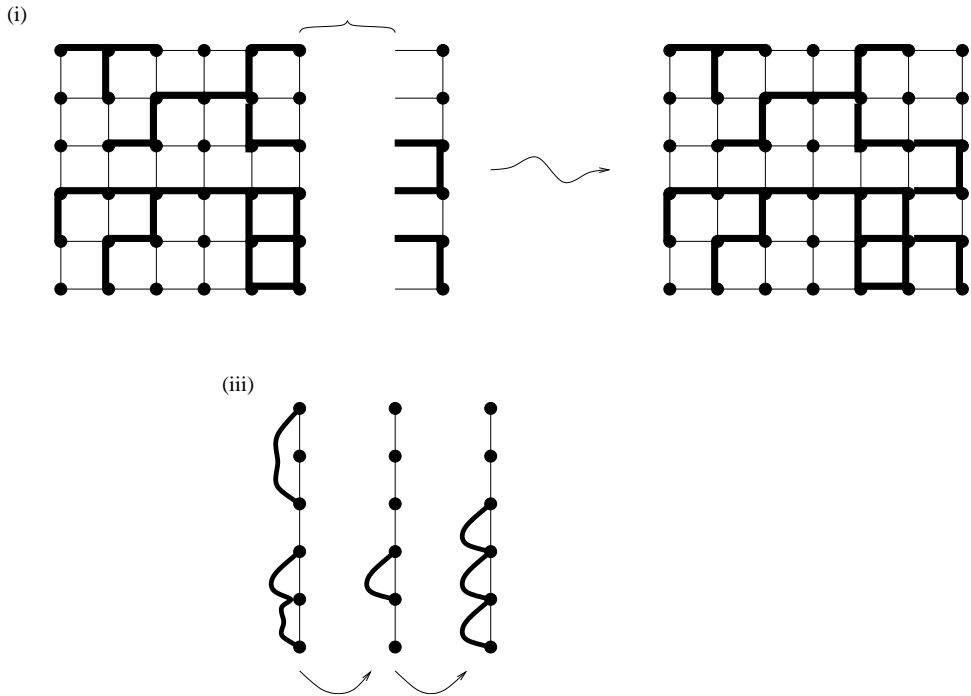}
\caption{\label{lattice2d2} (i) A subgraph of a square lattice and an
  extra layer. 
(ii) The corresponding new subgraph. 
(iii) A sequence showing: 
  the connectivity of the original subgraph (running $\#=12$); 
the connectivity
  after adding the new horizontal edges (running $\#=12+3$); 
  the connectivity after adding the new vertical edges 
  (running $\#=12+3-2$).}
\end{figure}

The physical setting 
for the \TL\ algebra 
from which we want to generalise is well 
known 
\cite{Baxter73,BloteNightingale82,Martin91}. 
For $G$ a graph with vertex set $V_G$ and edge set $E_G$ 
we associate a $Q$-state Potts variable 
$\sigma_i \in \underline{Q} \; := \; \{1,2,..,Q \}$
to each $i \in V_G$. 
One starts with the Potts Hamiltonian for $G$
\[
H_G = \sum_{<ij> \in E_G} \delta_{\sigma_i, \sigma_j}
      +h \sum_{i \in V_G} \delta_{\sigma_i, 1}
\]
We take magnetic field parameter $h=0$ and form the partition function
\[
Z_G (\beta) \; = \; \sum_{\{\sigma_i \}} \exp( \beta H_G )
 \; = \; \sum_{\{\sigma_i \}} \prod_{<ij> \in E_G} 
                          \exp( \beta \delta_{\sigma_i, \sigma_j} )
 \; = \; \sum_{\{\sigma_i \}} \prod_{<ij> \in E_G} 
                          ( 1 + v \delta_{\sigma_i, \sigma_j} )
\]
where $v=\exp(\beta)-1$. 
Expanding we have
\eql(fred)
Z_G (\beta)
 \; = \; \sum_{\{\sigma_i \}} \sum_{G' \in \Power(E_G)} 
             \prod_{<ij> \in G'}  v \delta_{\sigma_i, \sigma_j}
 \; = \; \sum_{G' \in \Power(E_G)} 
               v^{|G'|} Q^{\#(G')}
\eq
where $|G'|$ is the number of edges and 
$\#(G')$ is the number of connected components of $G'$ regarded
as a subgraph of $G$ in the obvious way. 
Example:
Figure~\ref{lattice2d2}(i) shows a subgraph $G'$ on a square lattice,
with $\#(G') = 12$.


We can now consider the RHS of (\ref{fred}) 
in its own right (as a `dichromatic'
polynomial in variables $v$ and $Q$). The exercise is to construct a
transfer matrix formultation in which to compute it, in cases where
$G$ has (`time') translation symmetry.
We also require that $G$ embeds in some Euclidean space and that its
edges (and hence the Potts interactions) are local. 
However even this is not enough to make the interactions in the
dichromatic polynomial formulation local, since $\#(G')$ is not
local. 
Instead we need to introduce an entirely different state space.
Although the restriction is not necessary, for the sake of simplicity
we will describe this by considering the example of the 
$n$-site wide square lattice. 

In adding an extra layer to this lattice we are adding 
$2n-1$ edges. 
As ever in a transfer matrix formalism, 
the problem is to find a set of states 
which keep enough information about 
the old lattice $G$ to determine   $\#(G')$ for the new one. 
It will be evident that each state must record which of the last layer
of vertices in $G$ are connected to each other 
(by some route in $G$ --- cf. Figure~\ref{lattice2d2} (i), (ii) and (iii)). 
Neither the details of the connecting routes nor any other information
is needed, thus our state set is simply contained in the set of 
partitions of the last layer of vertices. 
It is straightforward to see that 
(in the square, or otherwise plane, lattice case)
precisely the set of `plane' partitions are needed.
These are the partitions realisable by noncrossing paths in the
interior when the vertices are arranged around the edge of a disk. 


\begin{figure}
\[
\includegraphics{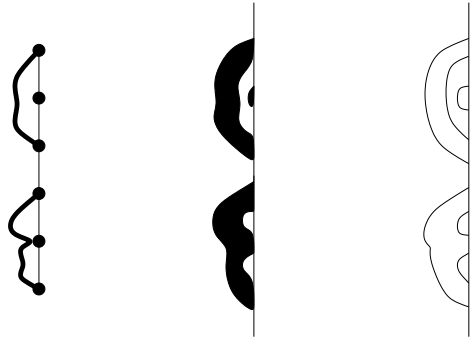}
\]
\caption{\label{lattice2d3} TL diagram.}
\end{figure}


Pictures of such paths are called Whitney diagrams \cite{Martin91}.
If instead we represent plane partitions by boundaries of connected
regions these diagrams become \TL\ (or boundary) diagrams
on the disk. Note that these are plane {\em pair} partitions
(of double the number of vertices). 
See Figure~\ref{lattice2d3} for an example. 
Note that the original lattice itself has all but disappeared from the
state space (replaced by a topological/combinatorial construct).


Finally we note that in order to compute correlation functions some
further information must be retained
(essentially the details of connections also with the vertices
on the left-hand side of the graph in Figure~\ref{lattice2d2}). 
This corresponds to \TL\
diagrams on the rectangle -- i.e. with both in-vertices and out-vertices. 
These diagrams may be composed by juxtaposition at one edge of the
rectangle when the number of states agrees. 
With an appropriate
reduction rule for interior loops (replace by a factor $\sqrt{Q}$)
this becomes the \TL\ algebra
(indeed category, indeed monoidal category). 


\begin{figure}
\[
\includegraphics{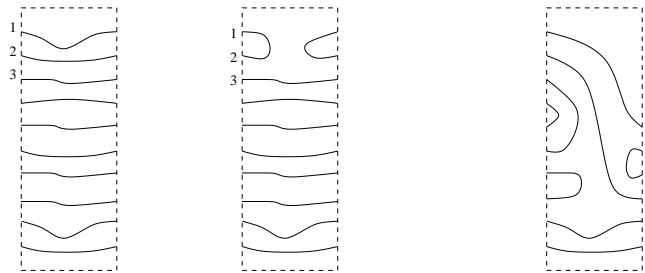}
\]
\caption{\label{TLgen} TL identity diagram, diagram $D_1$, and a
  diagram with different numbers of in and out-vertices.}
\end{figure}


NB, casting the state space in this form is certainly beautiful and
computationally convenient (see \cite{Martin2000}), but it is not the same as
integrability. Since the Potts model is integrable under certain
conditions solutions to the Yang--Baxter equations can be constructed
using \TL\ diagrams, but such exercises will not be our focus in the
present paper. 

The following set of \TL\ diagrams generate the \TL\ algebra
on $n$ vertices (i.e. $n$ in- and $n$ out-vertices). 
The identity diagram
is the rectangle in which each in-vertex is connected to the
corresponding out-vertex. The diagram $D_i$ is like the unit except
that in-vertices $i$ and $i+1$ are connected, and out-vertices $i$ and
$i+1$ are connected. (See Figure~\ref{TLgen}.)
The generators are $D_1,..,D_{n-1}$. 
As already noted, composition $B \circ C$ is by juxtaposition so 
that the out-vertices of $B$ meet the in-vertices of $C$ 
(becoming internal points in the new diagram).  
The state space we have constructed induces a representation $R$ of these
elements. The transfer matrix is then 
\[
T = \prod_{i}(1+\frac{v}{\sqrt{Q}} R(D_{2i}) 
     \prod_{i}(\frac{v}{\sqrt{Q}}+R(D_{2i-1}))
\]
and 
\[
Z(\beta ) = Tr(T^n)
\]

Finally,
the trace can be decomposed into the irreducible representations 
in $R$ (amongst other partial diagonalisations).
The close relationship this engenders between representation theory
and correlation functions (see e.g. \cite{Martin2000}) is 
what we aim to generalise. 

We require to generalise this picture in particular to higher dimensions.
For the sake of definiteness let us now consider the cubical lattice.
The direct generalisation of the Potts model 
leads us to certain graph \TL\ algebras --
subalgebras of the partition algebra \cite{Martin94}. 
However these are extremely intractable (see \cite{DasmahapatraMartin96}).
Here we take a different approach, staying closer to TL diagrams. 

Both mathematically and physically it is convenient to assemble the
\TL\ algebras into the \TL\ category
(diagrams with different numbers of in- and out-vertices
\cite{Martin91},
see later for details). 
Accordingly we approach the problem of generalisation here by 
casting the problem in a categorical framework, and 
defining a number of categories generalising this category. 
The claim is that this is natural, and ultimately makes the 
exposition more efficient. 

Hereafter we draw and compose 
diagrams from bottom to top, rather than from left
to right (i.e. subsequent pictures are rotated through $90^o$ compared
to those above). This is merely a space saving device. 


The \TL\ category 
$\CC{}_T = (\N, \hom_T(-,-), \circ)$ 
is monoidal \cite{Kassel95,Muger06} and generated as such by the diagrams 
\[ 
\includegraphics{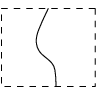}
\in \hom_T(1,1) 
\qquad
\includegraphics{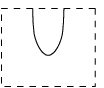}
\in \hom_T(2,0) 
\]
(cf. the diagrams above) and inversion. 
In this orientation it is the monoidal composition that is drawn from
left to right. 
For example, if $*$ is the monoidal composition then 
\[
\includegraphics{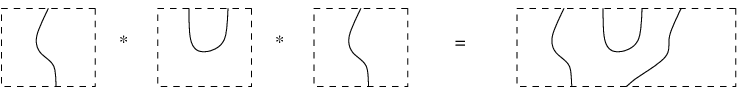}
\]
\label{ss:de:TLcat}
(Strictly speaking one works in the $\C$-linear category
$\CC{}_{T(\delta)}$ in which 
\eql(bubblego)
\raisebox{-10pt}{\includegraphics{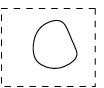}} 
\;\; \equiv \; \delta \; 
\raisebox{-10pt}{\includegraphics{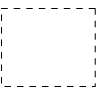}} \;\; \in \hom_T(0,0)
\eq 
and, by our loop replacement rule, 
$\delta=\sqrt{Q} \in \C$.)

As noted, these diagrams are shorthand for plane pair partitions.
However for the purposes of generalisation it will be convenient to 
begin (in Section~\ref{s:concr}) by treating them more literally
as lines embedded in $\R^2$. 

\medskip

One can view a curve properly embedded in a frame $F$ (an interval of
$\R^2$) as connecting two points on the boundary (if it is open), 
or as a separation of $F$ into two connected components. 
The Jordan curve theorem can be seen as a connection between these
views. As we see above, \TL\ diagrams are generated by such embeddings
(eventually one is only interested in the topology rather than the
specific embedding). 
In generalising to higher $d$ the two perspectives suggest distinct
generalisations. The Jordan theorem may be generalised to the Jordan--Brouwer
theorem \cite{Brouwer12,Wilder30}:
{\theo{ 
A closed (or properly embedded) $d-1$-manifold immersed in Euclidean
space $\R^d$ ($d>2$) separates $\R^d$ into two domains of which it is the
common boundary.
}\smallskip}

\noindent
And this gives the spirit of the approach we shall adopt here.
\\
Remark:
The converse (that a bounded and closed point set that 
separates, every point of which is accesible from each domain, 
is a manifold) is not true in
general, but our considerations here are limited by what can be built as a
limit of lattice objects (or by related 
physical considerations), so it is safe
enough heuristically 
to consider manifolds and separating sets as interchangable. 

\subsection{Overview}\label{S:over}
Our first task is to give a more precise description of the diagrams from which
we shall 
construct our generalisation. 

These diagrams, which we  call ``\CD s'', 
form a continuously infinite set. 
We require that our generalisations of TL algebras 
retain the property of finite-dimensionality.
Accordingly we will partition the continuously infinite set of 
concrete diagrams into a countably infinite collection of finite sets, each 
set 
characterised (in the three-dimensional case)
by a configuration of loops in the plane. The elements of 
these finite sets are the equivalence classes of concrete diagrams of an 
equivalence relation that we call ``heterotopy'', 
which extends ambient isotopy 
by allowing and regulating certain topology changes. 
The next task is to define a notion of 
composition for these equivalence classes based on the intuitive idea of 
concatenating concrete diagrams by pasting the bottom of one of them to the 
top of the other one. 
Then we can  define our category.
This category belongs to the beautiful class of diagram categories,
having a rich structure in representation theory, beside generalising
our physical setting. We conclude with some remarks on 
the structure of the category (although a detailed investigation
of its representation theory will be presented elsewhere). 

\subsection{Glossary of terms} \label{ss:gloss}
\newcommand{\pif}{\pi}
\newcommand{\IF}{I}


\newcommand{\gitem}[4]{ #1 & #2 & #4\ref{#3} \\ }

\begin{tabular}{llll}
\gitem{ $\bub(D)$}{ number of {\em bubbles} in \CD\ $D$}{s:concr}{\S}
\gitem{ $\CC{d}$}{category of \CD s in dimension $d$}{pr:cat1}{Prop.}
\gitem{ $\Ring\CC{d}(\delta)$}{$\Ring$-linearised $\CC{d}$ 
   quotient}{ss:category}{\S}
\gitem{ $\CC{}_{\Pa}$}{partition monoid base category}{ss:de:Pmcat}{\S}
\gitem{ $\CC{}_{\Pa(\delta)}$}{partition category
   with parameter $\delta\in\Ring$}{ss:de:Pmcat}{\S}
\gitem{ $\CC{}_{\she}$}{\shet\ category}{s:sh}{\S}
\gitem{ $\CC{}_{T(\delta)}$}{\TL\ category}{ss:de:TLcat}{\S}
\gitem{ $\DV_{\pm}D$ }{ upper/lower concrete boundary of $D$ }{de: DV}{Def.}
\gitem{ $E^d_t $}{ interval of Euclidean $d$-space}{ss:de:E}{\S}
\gitem{ $\genus(D)$}{ genus of three-dimensional \CD\ $D$}{S:handle}{\S}
\gitem{ $\IF_F$ }{`straight' isomorphism in $\pif_F$ }{de:pif}{Def.}
\gitem{ $\Pa(T)$ }{set of partitions of set $T$}{ss:de:Pmcat}{\S}
\gitem{ $\Power(T)$ }{power set of set $T$}{ss:gloss}{\S}
\gitem{ $\Pt_0{X}$ }{ point set of embedded manifold $X$}{s:concr}{\S}
\gitem{ $\Pt_1{X}$ }{ set of connected components of $X$}{s:concr}{\S}
\gitem{ $\conn(D)$ }{partition of connected components of $\partial D$}{de:con}{Def.}
\gitem{ $\pif_F$ }{ set of pre-isomorphisms in $\St{d}[F,F]$ }{de:pif}{Def.}
\gitem{ $\Rem(D)$ }{$D$ with bubbles removed}{s:concr}{\S}
\gitem{ $\Sph^d$ }{ $d$-sphere}{ss:gloss}{\S}
\gitem{ $\St{d}$ }{ set of \CD s in $d$ dims; lower hyperplane
  at $t=0$}{de:St}{Def.}
\gitem{ $\St{d}[F_+,F_-]$ }{ subset of $\St{d}$; 
  $D \in \St{d}[F_+,F_-]$ implies $\DV_{\pm} D = F_{\pm}$}{de:StFF}{Def.}
\gitem{ $\SD{d}$ }{ set of representatives of isotopy classes of \CD s }{de:repccd}{Def.}
\gitem{ $\SD{d}[F_+,F_-]$ }{ set of representatives of isotopy classes in
  $\St{d}[F_+,F_-]$}{de:repccd}{Def.} 
\gitem{ $\Sto{d} $}{ set of \CDD s of dimension $d$}{de:Sto}{Def.} 
\gitem{ $\Sto{d}(n) $}{ subset of \CDD s with $n$ components}{de:Sto}{Def.} 
\gitem{ $\Sd{d} $ }{ set of isotopy class representatives of $\Sto{d} $}{de:repbcd}{Def.}
\gitem{ $\Smin{d}$ }{ set of minimal \CD s}{de:Smin}{Def.}
\gitem{ $\Smin{d}[F_+,F_-]$ }{ subset $\Smin{d} \cap \St{d}[F_+,F_-]$ }{de:Smin}{Def.}
\gitem{ $\SSmin{3}[F,F']$ }{ set of heterotopy classes of diagrams in
  $\Smin{3}[F,F']$ }{de: het}{Def.}
\gitem{ $\SSmins{3}[F,F']$ }{ set of \shet\ classes of diagrams in
  $\Smin{3}[F,F']$ }{de: het}{Def.}
\end{tabular}



\section{Concrete diagrams}\label{s:concr}
Fix  $d \in \N$.
It is convenient to formulate our underlying space $\R^d$
with one prefered direction,   
called `time' $\tz$. 
Time totally orders spatial hyperplanes 
(hyperplanes perpendicular to the $\tz$--axis)
without further coordinatisation. 
Given a point $(v,\tz) \in \R^d$ we can associate a projection
$v \in \R^{d-1}$. 
For $\tz \in \R$ write $f_\tz$ for the embedding 
\eql(f proj)  f_\tz : \R^{d-1} \rightarrow \R^d \eq
\[ f_\tz(v) = (v,\tz) \]

As usual, if $M$ is a manifold with boundary we write $\partial M$ for
the boundary. Thus  $\partial M$ is a manifold without boundary, of
dimension one less than $M$. 
(There is a version of what follows for manifolds `with corners', but
we will not treat it here.)
For clarity, if we want to treat a manifold embedded in $\R^d$ as a
set of points we may write $\Pt_0(M) \subset \R^d$; 
while $\Pt_1(M)$ will denote
the partition of this point set into the connected components 
$\{ M_i \}$ of $M$. 
Write $|M|$ for the number of  connected components; 
$\bub(M)$ for the number with $\partial M_i = \emptyset$; and 
$\Rem(M)$ for the manifold obtained from $M$ by removing every $M_i$
with  $\partial M_i = \emptyset$. 

Recall that a manifold embedding $f:X \rightarrow Y$ is {\em proper} if 
\[
f ( \partial X ) =  f(X) \cap \partial Y 
\]
(for smooth manifolds there is also a transversality condition). 

\label{ss:de:E}
Fix $t \geq 0$. Then  
$$
E_t = E^d_t \; := \; \Re^{d-1} \times [0,t] .
$$ 
The subspaces $ \Re^{d-1} \times \{0\}$ 
and $ \Re^{d-1} \times \{ t \}$ are the 
components of the boundary of $E^d_t$. 



\begin{definition} \label{def:concrete}
A \CD\  in $E^d_t$ 
is a collection $D=\{D_i \}$ of connected  
compact submanifolds (`components')
of codimension-1, 
properly embedded in $E^d_t$, such that 

$(i)$ the submanifolds do not intersect;
 
$(ii)$ each boundary component is topologically a $(d-2)$-sphere.

\label{de:St}
\noindent
Let $S^d_t$ denote the set of \CD s in $E^d_t$; and 
$S^d = \cup_t S^d_t$ --- the set of \CD s. 
\\
(Note that each $D_i$ is usually a $(d-1)$-dimensional manifold.
However we include the case $\St{d}_0$ as a limit, 
and in this case each $D_i$ is just a $(d-2)$-sphere.)
\end{definition} 

\begin{figure}
\eql(D=d=2ex)
D= \;\; \;\; 
\raisebox{-.95021in}{
\includegraphics{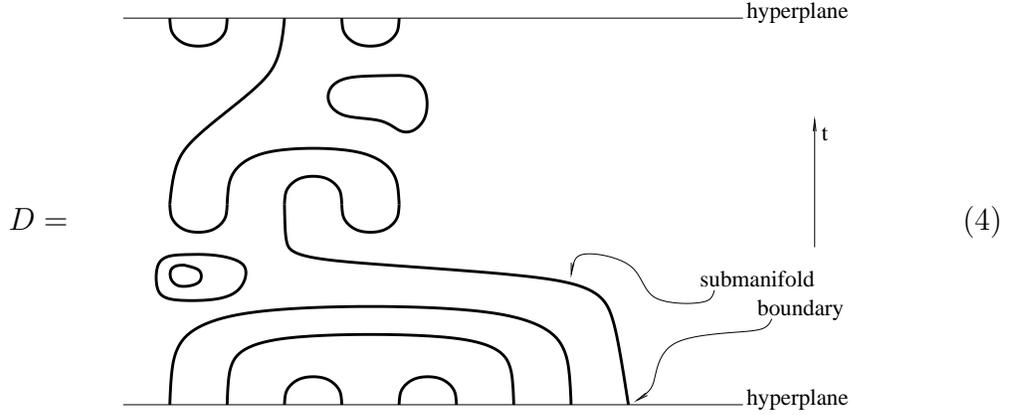}
}
\eq
\caption{\label{D=}
Example \CD\ in $d=2$ with $|D|=10$ and $b(D)=3$.
}
\end{figure}

\begin{figure}
\[
\epsfbox{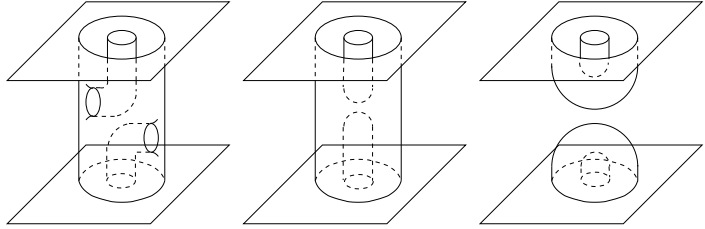}
\]
\caption{\label{fig(())} 
   A representation of examples of \CD s in $\reals^3$
   whose upper and lower 
\CBC s  
each consist of two concentric loops. }
\end{figure}

\begin{figure}
\eql(parti1)
D = \myfig{-.50}{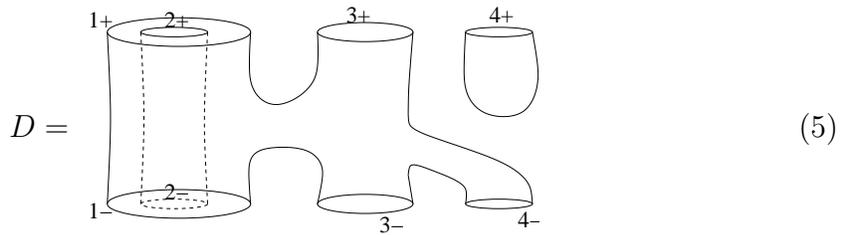}
\eq
\caption{\label{f:parti1} A representation of a \CD\ 
with labelled boundary components.}
\end{figure}

\noindent
There are examples in Figures~\ref{D=}, \ref{fig(())}, 
\ref{f:parti1} and~\ref{f:tworealisations}. 
The main property of each $D_i$, 
for our purposes, will be that it separates $E^d_t$ into
two regions, and is the boundary of both, with every point in it
accessible from both. 

\de{
Let $E$ be a Euclidean space and $D$ a subset. 
Then $C_E(D)$ is the number of connected components of $E \setminus D$.
}


\xfutnote{
Interval $\Uni(D)\; := \; \Re^{d-1} \times [t_1,t_2]$, 
is called the {\em universe} of $D$.
}

It is assumed that all the components of all concrete diagrams lie
within some finite interval of $\R^d$. 
So we assume in particular that there is a region of $\R^d$ (or even $E^d_t$) 
spatially very far away from 
the components of any diagram. 
(This `outer' region is connected unless $d=2$. 
In $d=2$ we will consider the outer region to be the part on the right.)

\xfutnote{
The $\tz_2$ (\resp\ $\tz_1$) boundary hyperplane   
shall be called, and drawn as, the {\it upper (lower) hyperplane}. 
}
The set of boundary components contained in the upper 
(\resp, lower) hyperplane is called the 
{\it upper (lower) boundary configuration}. 


\de{
A \CDD\ in dimension $d$  is a collection of non-intersecting 
topological $(d-1)$-spheres (i.e. without boundary) embedded in $\R^d$.
}
Thus a \CBC\ of a \CD\ in dimension $d$ is a \CDD\ in dimension $d-1$.


In $d=2$ the hyperplanes of a \CD\ are simply two parallel 
lines (the {\it edges} of the diagram). 
The 
components are one-dimensional submanifolds embedded between the two edges. 
If a component has a boundary, it consists of exactly two distinct 
0-spheres (points) which may be both on the same edge, or one on each 
edge. 
Those components without a boundary are homeomorphic to closed loops. 
The upper (lower) boundary configuration is the set of boundary points in 
the upper (lower) edge. 
Thus we see that those concrete diagrams in 
$\reals^2$ with equal numbers of boundary points in both edges 
are concrete \TL\ diagrams. 

In $d=3$ the hyperplanes are two parallel planes and the components are
essentially Riemann surfaces with (possibly empty) boundaries attached to the limiting 
planes. See Figure~\ref{fig(())}. The boundaries contained in a limiting plane define an 
arrangement of non-intersecting closed loops in that plane which we call 
a {\it \CLC}.

\subsection{
Categories of \CD s} \label{ss:category}

\label{de:Sto}
Let $\Sto{d} $ denote the set of   
\CDD s in dimension $d$.
That is
 $$
\Sto{d} = S^{d+1}_0
$$ 
Let $\Sto{d}(n) $ be the subset of \CDD s with $n$ components.


\de{ \label{de: DV}
Let $D \in \St{d}$. 
Then $\DV_{\pm}D$ is its upper
($+$) or lower ($-$) concrete boundary configuration 
(that is, a collection of non-intersecting topological $(d-2)$--spheres
embedded in $\R^{d-1}$).  
}
We adopt the convention that  $\DV_{\pm}D$ does {\em not} record the
$t$-coordinate
of its copy of $\R^{d-1}$ as a hyperplane in $\R^d$. 
Thus for $D \in \St{d}$ we have $\DV_+ D \in \Sto{d-1}$.  

\noindent
Example: in $d=3$, $\Pt_1{\DV_+ D}$ is  
a set of loops.


\xfutnote{{
\de{ \label{de:St}
Let $\St{d}$ be the set of concrete diagrams in $d$ dimensions 
whose lower hyperplane is at time $t=0$.}

Thus every concrete diagram is a time translate of some 
$D \in \St{d}$. 
That is, the set of  arbitrary concrete diagrams may be
partitioned into time-translate classes,  
with $\St{d}$ being a set of representative elements. 
}}

\de{ \label{de:StFF}
For $F_{\pm} \in \Sto{d-1}$, let $\St{d}[F_+,F_-]$ be the subset 
of $\St{d}$ such that
  $D \in \St{d}[F_+,F_-]$ implies $\DV_{\pm} D = F_{\pm}$. 
Let  $\St{d}[F_+,F_-](n)$ be the subset of 
$\St{d}[F_+,F_-]$ with $n$ components. 
Define $\St{d}_{\Rem}[F_+,F_-] = \Rem(\St{d}[F_+,F_-])$
similarly.
}

\begin{figure}
\[
 \myfig{-.50}{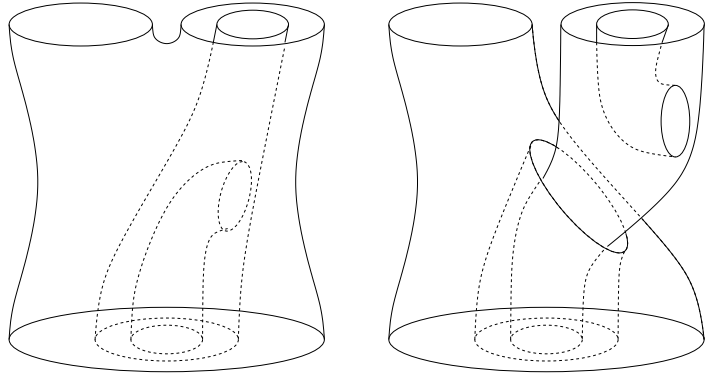}
\]
\caption{\label{f:tworealisations} Representations of \CD s 
in $\St{3}[F,F'](2)$ with 
$[F]_{\ii} = ()(())$ and $[F']_{\ii}=((()))$
(see (\ref{iclass}) for $\ii$-class notation). 
}
\end{figure}

\noindent See Figure~\ref{f:tworealisations} for examples. 

\begin{figure} 
\epsfbox{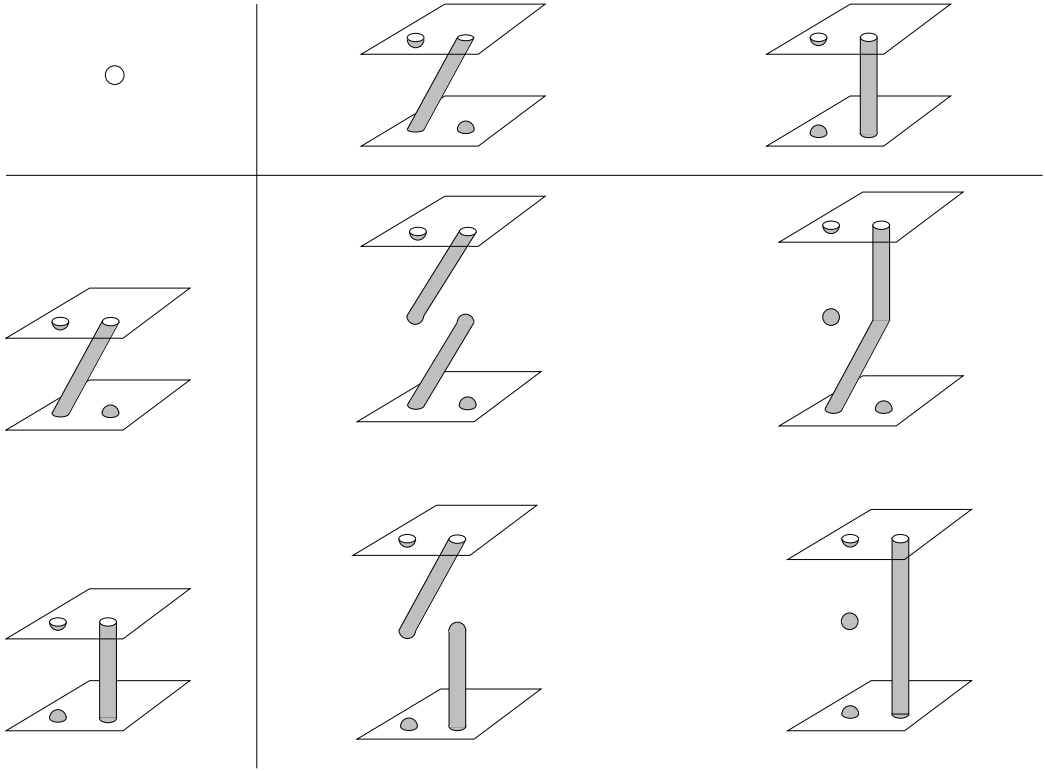}
\caption{\label{iso2} 
   A representation of examples of composition of concrete diagrams in $\R^3$.
}
\end{figure}

Let $A \in \St{d}_{t_1}$, 
$B \in \St{d}_{t_2}$ be such that $\DV_- A = \DV_+ B$ 
(i.e. $A \in \St{d}[F_+,F]$,  $B \in \St{d}[F,F_-]$ for 
some $F_+,F,F_-$).
Write $A \circ B$ for the point set that
 coincides
with $B$ as a point set below and on the upper hyperplane  
$t_2$ of $B$, 
and with the translate of $A$ above and on $t_2$. 
(Examples are shown in Figure~\ref{iso2}.)

\pr{\label{cpmpos}
Set  $A \circ B$ is itself the point set of a \CD
\footnote{Remark: the composite embedded manifold will only be smooth
  if the factors are smooth,
and not necessarily smooth if the factors are not transversal, 
but this need not concern us.
}
in $\St{d}_{t_1+t_2}$.
}

We now identify  $A \circ B$ with this diagram, and hence define
\[ 
\circ: \St{d}[F_+,F]
         \times \St{d}[F,F_-]
  \rightarrow \St{d}[F_+,F_-]
\]
\eql(AoB)
(A,B) \mapsto A \circ B 
\eq


Consider the triple 
$$
\CC{d} = (\Sto{d-1}, \St{d}[-,-],\circ)
$$ 
consisting of `object' set the set of 
concrete boundary configurations; 
and for each pair of objects $E,F \in  \Sto{d-1}$ 
the collection of `morphisms' ${\rm Hom}_{\CC{d}}(E,F)$ 
$=\St{d}[E,F]$;
and the composition defined above.
\pr{ \label{pr:cat1}
The triple $\CC{d}$ is a category. 
}
{\em Proof:}
Since the construction implied for $(A_1 \circ A_2) \circ A_3$
requires disection of the same \CD\ as  
 $A_1 \circ (A_2 \circ A_3)$ it follows that $\circ$ is associative. 
Each ${\rm Hom}(F,F)$ has a unit (the \CD\ of duration zero). 
\Qed


\newcommand{\dcirc}{\bullet}

If $\Ring$ is a ring and  $C$ a set (or category) write $\Ring C$ for
the free $\Ring$-module with basis $C$ 
(\resp\ the $\Ring$-linear category extending $C$,
for which the hom-set 
$\hom_C(F,G)$ is a basis for $\hom_{\Ring C}(F,G)$).

By a mild abuse of notation we call a complete collection of hom-space
bases for a $\Ring$-linear category $C$ a basis for $C$. 
A basis is {\em categorical} if it forms a subcategory. 

For $\delta \in \Ring$ we define a category 
$$
\Ring\CC{d}(\delta) = (\Sto{d-1}, \Ring\St{d}_{\Rem}[-,-],\dcirc)
$$ 
by 
$A \dcirc B = \delta^{b(A \circ B)} \Rem( A \circ B )$
on the basis
(it is straightforward to check associativity as before). 
Note that $\CC{d}(1) = (\Sto{d-1}, \St{d}_{\Rem}[-,-],\dcirc)$
is a subcategory (with $\delta=1$ making $\dcirc$ close on
concrete diagrams in an obvious way). 

\subsection{Functors to partition categories}
\label{ss:de:Pmcat}

\newcommand{\ddotcup}{\stackrel{\bullet}{\cup}}
\newcommand{\und}[1]{\underline{#1}}

For $T$ a set let $\Pa(T)$ denote the set of partitions of $T$.
For $p \in \Pa(T)$ we may write $s \sim_p t$ if $s,t$ in the same part
in $p$ (acknowledging the natural bijection between partitions and equivalence
relations). We may also write $[s]_p$ for the part (or equivalence
class) of $p$ containing $s$.  
For $n \in \N$ define $\und{n}=\{1,2,...,n \}$. 

The forced disjoint union of two sets is 
$A \ddotcup B = \{ (a,1), (b,0) | a \in A, b\in B \}$,
or for $i \in \Z$, 
$A \ddotcup_i B = \{ (a,i+1), (b,i) | a \in A, b\in B \}$. 
Let $\rho$ be any relation from any set $T$ to $\Re$.
Then for each $i \in \Re$ define a new relation $\upsilon_i (\rho)$ 
from $T$ to $\Re$ by $(s,x) \mapsto (s,x+i)$.
For example, $\upsilon_i (A \ddotcup_{j} B )  
= A \ddotcup_{i+j} B$. 

For $F_{\pm} \in \Sto{d-1}$ in particular then by $F_{+} \ddotcup F_{-}$ 
we will intend the union of the sets $\Pt_1(F_{\pm})$ of connected
components.
\\ 
For example
 the concrete diagram
$D$ in Figure~\ref{f:parti1} has 
\[
\DV_+ D \ddotcup \DV_- D = \{ 1+, 2+, 3+, 4+, 1-, 2-, 3-, 4- \}
\]
in a notation in which $(1,1/0) \mapsto 1\pm$ and so on. 

\de{ \label{de:con}
Fix $d$. For each pair $F_+,F_- \in \Sto{d-1}$ define 
\[
\conn : \St{d}[F_+,F_-] \rightarrow \Pa( F_+ \ddotcup F_- )
\]
as follows. 
Let $D$ be a \CD.
The {\it connectivity} $\conn(D)$   
is the partition of 
$\DV_+ D \ddotcup \DV_- D$
each element of which is the set
of all topological $(d-2)$-spheres
(loops in $d=3$) bounding a single component in $D$. 
}


\noindent
Examples: In $d=2$ every $p(D)$ is a partition of the set of boundary
points into pairs. 
The case in equation~(\ref{D=d=2ex}) has 
$p(D)=\{\{1+,2+\},\{3+,9-\},\{4+,5+\},...\}$ 
(in the natural labelling scheme). 
\\
In $d=3$ the concrete diagram
$D$ in Figure~\ref{f:parti1},
with boundary loops labelled as indicated, 
gives 
$$
p(D) = \{ \{ 1+,1-,3+,3-,4- \},\{ 2+,2- \}, \{4+ \}  \}
$$


\de{
(1) A congruence relation $I$ on a category $\CC{}$ is an equivalence
relation on each hom set such that $f' \in [f]_I$ and $g' \in [g]_I$
implies $f' \circ g' \in [f \circ g]_I$ whenever the latter exists.
\\
(2) The quotient category $\CC{}/I$ has the same object class as $\CC{}$ but 
$hom_{\CC{}/I}(F,G) = hom_{\CC{}}(F,G)/I$ with the obvious composition
well-defined by congruence. 
}

\begin{figure}
\[
\includegraphics{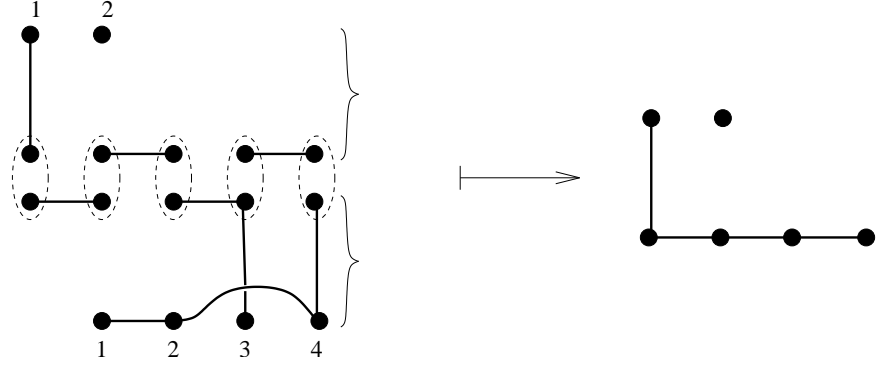}
\]
\caption{\label{partitioncomp1} Example of composition in the 
partition category. }
\end{figure}

The partition-monoid base category  
$$
\CC{}_{\Pa} = (S_{Fin}, \Pa(-\ddotcup -),\pcirc)
$$
has object class the class $S_{Fin}$ of all finite sets.
The composition
\begin{eqnarray*}
* : \Pa(S \ddotcup T) \times \Pa(T \ddotcup U) & \rightarrow & 
\Pa(S\ddotcup  U)
\\
(a,b) & \mapsto & a*b
\end{eqnarray*}
has $a*b$ given as follows (see Figure~\ref{partitioncomp1} for an example). 
With $s,s'  \in S\ddotcup U$ we have $s \sim_{a*b} s'$ if 
there is a sequence $s_0,s_1,s_2,...,s_k$ with 
$s_0 = s$ (case $s \in \emptyset\ddotcup U$) 
or  $s_0=\upsilon_1(s)$  (case $s \in S\ddotcup\emptyset $) 
and 
$s_k = s' \mbox{ or } s_k=\upsilon_1(s')$;
such that either  
$s_{2i} \sim_{b} s_{2i+1}$ 
and 
$s_{2i+1} \sim_{\upsilon_1(a)} s_{2i+2}$,
or 
$s_{2i} \sim_{\upsilon_1(a)} s_{2i+1}$  
and
$s_{2i+1} \sim_{b} s_{2i+2}$. 
(See \cite{Martin94,Martin07} for a gentler introduction).   
\ex{
See Figure~\ref{partitioncomp1}.
The upper bracketed part in the figure represents an element $a$ of
$hom(\und{2},\und{5})=\Pa(\und2 \ddotcup \und5 )$; and
the lower part is $b$ in $hom(\und{5},\und{4})$. 
The composite is the partition in $hom(\und2 ,\und4 )$ 
of the upper and lower rows of the composite diagram (ignoring
the middle groupings except in so far as the marked identifications
engender connections). Note also that over/under information is irrelevant.
\\
Specifically we have $(3,0) \sim_b (3,1)$, 
$(3,0) \sim_a (2,0)$, $(1,1) \sim_b (2,1)$ 
and so on; so
\[
(3,0) \sim_b (3,1) \sim_{v_1(a)} (2,1) \sim_{b} (1,1) \sim_{v_1(a)} (1,2)
\]
giving $(3,0) \sim_{a*b} (1,1)$, and so on.
}
We may regard  $\Sto{d-1}$ as a subset of  $S_{Fin}$ by 
regarding each object $F$  as a set of components,
rather than a point set.
Whereupon we may define $\CC{d}_{\Pa}$ as the full subcategory of
$\CC{}_{\Pa}$ with object set  $\Sto{d-1}$, and  

\newcommand{\FF}{{\mathcal F}} 

{\pr{\label{claim xcon}
The inclusion of  $\Sto{d-1}$ as a subset of  $S_{Fin}$ 
together with the assignment of partition $\conn(D)$ to each \CD\ $D$
defines a functor
\[
\FF:  {\mathcal C}^d \rightarrow {\mathcal C}_{\Pa}
\]
Specifically, 
if $A,B$ composable in the category ${\mathcal C}^d$  
then $\conn(A \circ B) = \conn(A) \pcirc \conn(B)$. 
}}
\noindent
{\em Proof:}
Let $A \in \St{d}[E,F]$ and $B\in \St{d}[F,G]$. Then 
$A \circ B \in \St{d}[E,G]$ so both partitions are of $E \ddotcup G$.
But there is a route (a path within a component) from 
boundary component $l$ to $l'$ in
$A\circ B$ iff there is a chain $l=l_1, l_2, l_3, ..., l_m=l'$ such
that there exist routes from $l_{2i-1}$ to $l_{2i}$ in $A$ and  
from $l_{2i}$ to $l_{2i+1}$ in $B$
(or similarly with the roles of $A,B$ reversed).
Map $\conn$ thus defines a congruence relation on $\CC{d}$
leading to $\CC{d}_{\Pa}$ as quotient.
\Qed


\noindent
Extending map $p$ $\Ring$-linearly, $\FF$
extends in an obvious way to a functor 
$$
\FF: \Ring\CC{d}(\delta) \rightarrow \Ring\CC{}_{\Pa(\delta)}
$$
where $\Ring\CC{}_{\Pa(\delta)}$ is 
the usual partition category \cite{Martin94}
generalising $\Ring\CC{}_{T(\delta)}$. 


\de{
A category is {\em finite} if every hom set is finite; or, if it is a
$K$-linear category, if every hom set has a finite basis.
}

Category $\CC{}_{\Pa}$ is finite, so the image category  
$\FF(\CC{d})$ is finite. 
This is certainly an interesting object for study,
both from the view of generalisations to Section~\ref{S:dichro}, 
and mathematically. 
The image category 
$\FF(\CC{2})$ is a kind of \TL\ category (like the ordinary TL category
$\CC{}_T$ 
except that the object set is $\Sto{1}$ instead of $\N$). 
However this is not the only way of thinking of the TL category, 
so before studying $\FF(\CC{d})$ we consider the construction using
{\em isotopy} of diagrams. 


\section{Isotopy and minimality of \CD s}


\begin{definition}
A \CD\ in $S^d$ will be called {\em minimal} if 
all its components are $(d-1)$-spheres with non-empty boundaries. 
The set of all minimal \CD s in $\St{d}[F,F']$ is written 
$\Smin{d}[F,F']$. 
\label{def:minimal}
\label{de:Smin}
\end{definition}

\noindent Examples:  
The \CD s in Figures~\ref{fig(())},\ref{f:parti1},\ref{f:tworealisations} 
(but not \ref{D=}) are minimal. 


One way to characterise the diagram basis of the \TL\ algebra is as
the set of equivalence classes of 
minimal 
\CD s in $d=2$ under {\em isotopy} \cite{Moise77}.
In what follows we show that this construction 
does not generalise automatically to
higher $d$, and provide a way to resolve this anomaly. 

We begin by recalling and extending the definition 
\cite[Ch.11]{Moise77} of isotopy.


\subsection{Isotopy and strong isotopy}
\newcommand{\sst}{s} 

\begin{definition} \label{def:isotopy}
{\rm (i)} 
By an {\it isotopy} on $\reals^d$ we shall mean a one-parameter family 
$j_\sst$, $\sst \in[0,1]$, 
of homeomorphisms of $\reals^d$ such that 
$j_{\sst}(x)$ is continuous in $\sst$ and $x$ and $j_0$ is the identity 
homeomorphism. 
\\
{\rm (ii)}
If $A$ and $B$ are concrete diagrams 
we say that $A$ is {\it isotopic} to $B$ 
($A \simi B$, $A \in [B]_{\ii}$)
if there is an isotopy $j_{\sst}$ of $\reals^d$ such that:

\medskip
{\bf I1} $j_\sst ( \Pt_{0}(A) )$ is the point set of a concrete diagram in 
$\R^d$ for all $\sst$; 

{\bf I2} $j_1( \Pt_{0}(A) ) = \Pt_{0}(B)$.
\medskip
\newline
We may write just $j(A)$ for \CD\ $B$ here 
(and $A=j^{-1}(B)$).
Thus each $j$ defines a map $j: \St{d} \rightarrow \St{d}$.
\\
We define isotopy of \CDD s similarly.
\\
{\rm (iii)}
For $A,B \in \St{d}[F,F']$, we say that $A$ is {\it strongly isotopic}
to $B$ if $A\sim_\ii B$ by an isotopy $j_\sst$ such that 
$j_\sst[\Pt_0(A)]$ is the point
set of a concrete diagram in $\St{d}[F,F']$ for all $\sst$. 
If $A$ is strongly isotopic to $B$ we write $A\sim_{\si}B$
or $A \in [B]_{\si}$.
\label{def:strongiso}
\end{definition}


\ex{
The following \CD s are all isotopic in $d=2$. 
The first two are strongly isotopic.
\[
\myfig{0}{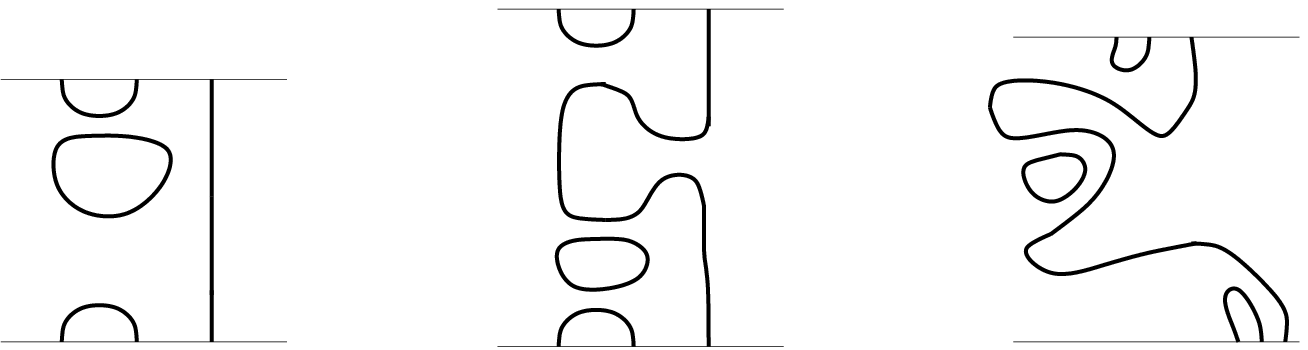}
\]
As noted above, these may all be regarded as representations of the
same TL diagram.
}

The two factor diagrams used in Figure~\ref{iso2} are isotopic but not
strongly isotopic.

\pr{ The relations $\simi$, $\simsi$ are equivalence relations. 
Bubble number $b(D)$ is an invariant of both. \Qed}
A significant difference between $\ii$ and $\si$ is the following.

\pr{
{\rm (i)}
For $D \in \St{d}[F,F']$ we have $[D]_{\si} \subset \St{d}[F,F']$.
Let $D_1 \in \St{d}[F,F']$ and  $D_2 \in \St{d}[F',F'']$,
and let $D_j' \simsi D_j$ ($j=1,2$). 
Then $D_1' \circ D_2' \simsi D_1 \circ D_2$ and we have a well-defined
composition on $\si$-classes. 
Thus the triple
$$
\CC{d}_{\si} = (\Sto{d}, \St{d}[-,-]/\si , \circ)
$$ 
is a quotient category of $\CC{d}$. 
\\
{\rm (ii)} If two concrete diagrams are strongly isotopic then their
connectivities are equal. Thus the functor $\FF$ factors through 
$\CC{d}_{\si}$. That is 
\[
\xymatrix{
\CC{d} \ar[rr]^{\FF} \ar[dr]^{I_{\si}} & & \CC{}_{\Pa} \\
& \CC{d}_{\si} \ar[ur]_{\FF_{\si}}
}
\]
commutes
(with $I_{\si}$ denoting the $\si$-congruence).
\Qed
}

\noindent
Equal connectivity does not
imply isotopy. For example, see Figure~\ref{stronghet1}. 
Thus $\CC{d}_{\si}$ is not a finite category.

\subsection{Hyperplane isotopy}

Note that I1 
of Definition~\ref{def:isotopy}
implies that points in a boundary remain cohyperplanar 
through the continuous transition realising an isotopy. 
The upper hyperplane may move bodily to a different time 
as we transform between isotopic \CD s, 
but it can be followed through the transformation.
(Although the limit in which the upper and lower hyperplane coincide
is allowed, where this makes sense.)
In this sense 
an isotopy on $\R^d$ {\em restricts} to a transformation on
a boundary hyperplane, which transforms between isotopic \CBC s.


We want to address the question of how to define a smaller category from
$\CC{d}$ by replacing \CD s with their $\ii$-classes,
as works for $d=2$. 
Two possible ways to go are: 
(1) try to find candidates for hom classes in  $\St{d}/\simi$ 
(note that this automatically implies a reduced object set);
(2) try to use the $\St{d}[F,F']/\simi$ as hom classes.
\\
We shall see that neither of these works directly unless $d=2$. 
Shortly we shall study the quotient sets $\St{d}/\simi$ and 
$\St{d}[F,F']/\simi$. We will need some preparations to deal with the
differences between $d=2$ and $d>2$. 


\pr{ \label{pr: tree}
(1) 
The set $\Sto{1}/\simi$ may be indexed by the  
natural numbers.
\footnote{(NB, this statement can be brought into line with the
sequel by a strict interpretation of the 0-sphere, but this need not
concern us here --- see \cite{AlvarezMartin07b}.)}
\\
(2)
For $d>2$ the set $\Sto{d-1}/\simi$ may be indexed by the set of
rooted trees 
(as in Definition~\ref{de:rooted} or, e.g., in \cite{NakanoUno03,Wilf94}). 
An isotopy class $[F]_{\ii}$ of elements of  $\Sto{d-1} $
can be indicated, non-uniquely, by a bracket notation like 
\eql(iclass)
[F]_{\ii} \; \mapsto \;\;  ()(()).
\eq
}
{\em Proof:} (1) The set $\Sto{1}$ is the set of boundary
configurations of $\St{2}$, i.e. points on the line. 
The intrinsic left-to-right order on a 
boundary point configuration 
is preserved by isotopy,
even though the precise location of points is not in general. 
As such there is precisely one {\em isotopy class} of point 
configurations for each number of points.
\footnote{It is the boundary class in $d=2$ which is fixed in 
 constructing the diagrams for a 
 specific \TL\ algebra (see later). 
 Hence there is one \TL\ algebra for each number of points.} 
\\
(2) For $d>2$
every sphere partitions $\R^d$ into its interior, its exterior, and
the intersection of their closures.
{\em Geometrical duality} 
\cite{Savit80}
places this partition into correspondence
with a graph consisting of a point for each open component and an edge
between them for the separating sphere.
\Qed



Recall that 
 $\Pt_1{D}$ is  a partition of $\Pt_0{D}$ into components.
However,  in writing examples we will usually use number labels
for the components, chosen for local convenience
(as in (\ref{parti1})).


In as much as an isotopy $j_\sst$ is continuous
we may consider following a particular component 
through the transformation.
(That is, 
while the point set of a component will change in general under 
a homeomorphism, we can consider the component's number label
to travel with its homeomorphic image.)
In this sense the isotopy 
may be considered to move the component around. 
Consider an isotopy taking a \CD\ to {\em itself} (via some homeomorphisms).
In the above sense such an isotopy permutes the components.
In particular the homeomorphic image of a component 
may not be the same component 
(although in $d=1$ this permutation is necessarily trivial,
by the nonintersection condition). 

\begin{figure}
\[
\myfig{0}{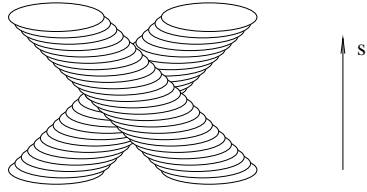}
\]
\caption{\label{homeo1} 
Depiction of a nontrivially permuting self-isotopy $j_\sst$ in $d=2$.}
\end{figure}

Consider a \CDD\ $F \in \Sto{2}$
consisting of two loops arranged as ()(). 
The example in Figure~\ref{homeo1} illustrates 
(discretely) a continuum of homeomorphisms $j_\sst$ 
realising a {\em self-isotopy} of this \CDD. 
This isotopy realises a nontrivial permutation.

\pr{\label{pr: Dj}
If $F$ is a \CDD, and $j_s$ an isotopy, in dimension $d-1$ then
(using $f_\tz$ from (\ref{f proj}))
\eql(DDDjF) 
\DDD{j_s}{F} 
:= \cup_{\tz \in [0,1]} f_\tz ( j_t (\Pt_0(F))  )
\eq
is (the point set of) a \CD\ in $\St{d}[F,F']$, 
with $F' = j_1(F)$.
}
To see this consider Figure~\ref{homeo1} as a single diagram in $d=3$.
\Qed


\de{ \label{de:pif}
Write $\pif_F$ for the set of \CD s in $\St{d}[F,F]$ 
arising as in equation~(\ref{DDDjF}). In particular 
$\IF_F \in \pif_F$ is given by $\IF_F = \DDD{j}{F}$ 
(as in equation~(\ref{DDDjF}))
in case $j_u$ is the identity homeomorphism for all $u$.
\label{def:unitf}
}



\pr{ \label{pr:skel1}
In $\CC{}_{\Pa}$ there is an isomorphism in $\hom(F,F')$ iff
$|F|=|F'|$. That is,  $\CC{}_{\Pa}$ has a skeleton with object set
$\N$. 
\\
In $\FF(\CC{d})$ there is an isomorphism in $\hom(F,F')$ iff
$F \simi F'$. That is,  $\FF(\CC{d})$ has a skeleton with object set
$\N$ (case $d=2$); or the set of rooted trees ($d>2$). 
}
{\em Proof:}
The results on  $\CC{}_{\Pa}$ are elementary. 
For  $\FF(\CC{d})$ note that the construction $\DDD{j_s}{F}$ {\em is}
an isomorphism. Then apply Proposition~\ref{pr: tree}. 
\Qed


\subsection{Representatives of isotopy classes}

\de{\label{de:repbcd}
Let $\Sd{d} \subset \Sto{d-1}$ 
be any complete set of representatives of classes 
in $\Sto{d-1}/\simi$ 
(one per orbit). 
}
Although there is much choice in the preparation of $\Sd{d}$, 
some restriction is convenient. 
\pr{
Within each orbit of $\Sto{d-1}/\simi$
there are representatives in which each component
  is a perfect $d-2$-sphere; and the centres of all these spheres are
  colinear (let us say, along the $x$-axis); 
and the intersections of the components with this
  line are  spaced along the line at unit intervals,
with the first intersection at $x=0$.
We may arrange the spheres so that the smallest appear first (reading
left to right) and then that those of equal size are ordered 
by heaviness (as in Definition~\ref{heavy} in Appendix~\ref{B:tree}). 
\Qed}
(Note that for $d>2$ the bracket notation of 
equation~(\ref{iclass}) 
may be used here
to indicate specific representatives.)

We will consider to be fixed for each $F \in \Sd{d}$ an enumeration of 
its components,  
to simplify labelling later on. 
(It does not matter which enumeration is fixed.
If we use the $x$-axial representatives above we can number in order
of first intersection of each component with the axis. 
Thus, for example, $()_1()_2((()_5)_4)_3$.) 

\de{\label{de:repccd}
Given $\Sd{d}$, then 
$\SD{d}$ is any set of representatives of classes of concrete diagrams under
  isotopy (one per orbit), such that 
$D \in \SD{d}$ implies $\DV_{\pm} \in \Sd{d}$. \\
For $F_{\pm} \in \Sd{d}$, $\SD{d}[F_+,F_-]$ is the subset of $\SD{d}$ such that
  $D \in \SD{d}[F_+,F_-]$ implies $\DV_{\pm} D = F_{\pm}$. 
}

\pr{ Every $D \in \SD{d}$ is in some $\SD{d}[F_+,F_-]$, thus
\[
\SD{d} = \cup_{F_{\pm} \in \Sd{d}}  \SD{d}[F_+,F_-]
\]
is a partition of $\SD{d}$. \Qed
}

\subsection{Handles and minimality}
\begin{figure}
\includegraphics{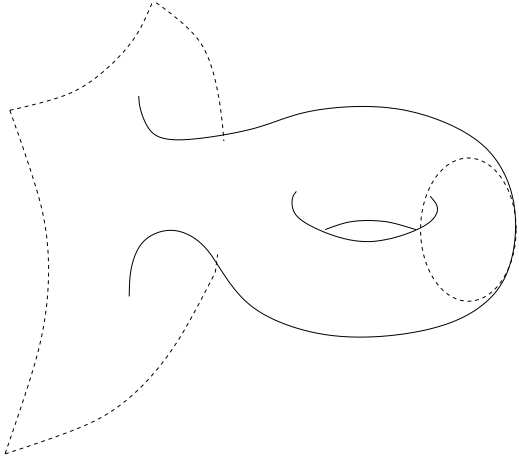}
\caption{\label{handle1} 
A representation of part of a \CD .
The closed path indicated by the dashed loop
on the right is a handle.}
\end{figure}

\label{S:handle}
A {\em handle} in $D \in \St{3}$ is a subset of $\Pt_0(D)$ 
that forms a closed path in a component, 
but can be removed without separating the component into two.
(Thus no minimal \CD\ has a handle.)
See Figure~\ref{handle1} for an example.
The genus $\genus(D)$ of $D$ is the maximum number of handles that can be
removed simultaneously without separating any component.  

\pr{\label{min iso}
Minimality, $b$ and $g$ are all $\ii$-invariants.
 \Qed }

The $\si$-classes in $\St{d}$ of elements in $\Smin{d}[F,F']$ lie in
$\Smin{d}[F,F']$.

\subsection{Class compositions: examples and counter examples}

First we look at $d=2$ and make contact with the TL category.
Then we will look at the new features when $d>2$. 

Noting Proposition~\ref{pr: tree}(1), for $m,n \in \N$ let us define
\eql(numcat)
\St{2}[m,n] = 
\bigcup_{F \in \Sto{1}(m); \; F' \in \Sto{1}(n)} \St{2}[F,F']
\eq
This argument-dependent notation allows us 
next to introduce certain triples, 
\\
$(\N, \St{2}[-,-]/\ii , \circ)$
and $\CC{2}_{\ii} = (\Sto{1}, \St{2}[-,-]/\ii , \circ)$
(that we shall show to be categories);
noting from the object sets 
that the hom sets differ, with the former to be understood
as defined by (\ref{numcat}). 
\pr{
(i) Subset $\St{2}[m,n]$ is a union of $\ii$-classes in $\St{2}$. 
\\
Let $D_1  \in \St{2}[F,F']$ and $D_2  \in \St{2}[F',F'']$.
\\
(ii) If  $D_j' \simi D_j$ ($j=1,2$) in $\St{2}$ then $D_1'$ and $D_2'$ are not
necessarily composable in  $\CC{2}$; but if they are then 
$D_1 \circ D_2 \simi D_1' \circ D_2'$. 
Thus we may define a composition $\circ$ using any 
such composable representatives which makes
$(\N, \St{2}[-,-]/\ii , \circ)$ a category. 
\\
(iii) If  $D_1' , D_1$ are \CD s in the same $\ii$-class in 
$\St{2}[F,F']/\ii$, and  $D_2' , D_2$ are \CD s in the same $\ii$-class in 
$\St{2}[F',F'']/\ii$, then 
$D_1 \circ D_2 \simi D_1' \circ D_2'$.
Thus we may define a composition $\circ$ which makes
$$
\CC{2}_{\ii} = (\Sto{1}, \St{2}[-,-]/\ii , \circ)
$$ 
a category. 
\\
(iv) On ignoring both the position and number of closed loops 
(so that $D \equiv D'$ if $\Rem(D)=\Rem(D')$)
the category in (ii) becomes the TL monoid category 
$\CC{}_{T(1)}$.
The category in (iii) contains an equivalent of this category as a skeleton.
}
{\em Proof:}
(i) Considering the isotopy class in $\St{2}$ 
of an element $D \in \St{2}[F,F']$ we
see that the class extends beyond $\St{2}[F,F']$ 
and contains elements in each  $\St{2}[E,E']$
with $E \simi F$ and $E' \simi F'$. 
By Proposition~\ref{pr: tree}(1) the appropriate sets are included in 
$\St{2}[m,n]$. 
\\
(ii)
Isotopy in $\St{2}$ thus induces an equivalence on 
\CBC s so that, cf. $\CC{2}$, the object set of any resultant category
is the set of equivalence classes. 
Again by Proposition~\ref{pr: tree}(1), the new object set is $\N$. 

In the plane, 
the isotopy class of $D \circ D'$ depends only on the isotopy
classes of $D$ and $D'$ (so long as $D,D'$ chosen composable).
\\
(iii) The same argument establishes a congruence relation in this
case.
\\
(iv) is straightforward.
The skeleton uses one object from each $\Sto{1}(m)$, $m \in \N$.
\Qed
\\
However, now we look at $d=3$:

Let $D_1,D_1',D_2,D_2' \in \Smin{3}[F,F]$.
It is easy to see that $D_j \simi D_j'$ ($j=1,2$) does not imply 
$D_1 \circ D_2 \simi D_1' \circ D_2'$ in general. 

\ex{Suppose $F$ is a concrete \CBC\ of form ()().
Let us label the two loops in the $t=0$ plane as 
\newcommand{\bba}{1-}%
\newcommand{\bbb}{2-}%
\newcommand{\bbba}{1+}%
\newcommand{\bbbb}{2+}%
$\bba,\bbb$, and their
translates in the upper plane as $\bbba,\bbbb$. 
Now consider any minimal concrete
diagrams $A,B \in \St{d}[F,F]$ whose components connect these loops as:
$p(A) = \{\{\bba,\bbba\},\{\bbb\},\{\bbbb\}\}$ and 
$p(B) = \{\{\bba,\bbbb\},\{\bbb\},\{\bbba\}\}$. 
Here $A \simi B$. Note, however, that $A\circ A \not\simi B\circ B$. 
This example is illustrated on the diagonal in Figure~\ref{iso2}.
}


\noindent
The above example shows us that, in $d=3$, 
composition $\circ$ does not pass to a
well defined composition on \CCDs. 
Thus { \CCDs} {\em per se} cannot have
quite the same standing in any algebra formulated in $d=3$ as they do
in $d=2$. 


This can be seen from a diagram category point of view.
The object set which $d>2$ diagrams can `factor through'
(in the sense of \cite{GreenMartin07}) cannot depend on an 
arbitrary numbering of components
(in $d=2$ it is not arbitrary -- distinct points on a line can be
naturally ordered). 

Our next objective is to make a generalisation of TL composition which
{\em does} work. We will need a suitable `plumbing kit', which we now
construct. 

\subsection{Pre-isomorphisms: symmetries of \CBC s}

Consider $F \in \Sd{d}$. 
Let $j_\sst$  
be any hyperplane isotopy which
fixes $\Pt_0{F}$ but permutes (possibly trivially)  $\Pt_1{F}$
(such as that illustrated in Figure~\ref{homeo1}). 
Let $\Sigma_{F}$ be the complete set of permutations of $\Pt_1{F}$
which can arise in this way.

\ex{Set $d=3$: Consider $F=\DV_+ D$ with $D$ as in (\ref{parti1}) above,
that is $\Pt_1{F} =(()_2)_1()_3()_4$. 
Then $\Sigma_F = \{ (), (34) \}$, where $()$ denotes
the trivial perm. 
In particular the trivial isotopy on $F$ achieves the trivial perm.
}

\pr{The set $\Sigma_{F}$ forms a subgroup of 
the symmetric group $\Sigma_{|\Pt_1{F}|}$ under composition of
permutations.}
{\em Proof:} 
It is enough to show closure. This can be seen by noting that 
self-isotopies can be composed (rescaling $\sst$).
\Qed


Suppose that $j_s$ achieves the permutation $\sigma \in \Sigma_F$.
Then recall that we may associate a \CD\  $\DDD{j_s}{F} \in \St{d}$ to it 
(from equation~(\ref{DDDjF})).

\pr{For given $F$ all \CD s of form $\DDD{j}{F}$ 
are isotopic
(i.e. irrespective of $j$ or the permutation $\sigma$ achieved).
All $\DDD{j}{F}$ are minimal.}
{\em Proof:} by construction. \Qed


We write $D_{\sigma}$ for a $\DDD{j}{F}$ 
whose $j$ achieves the permutation $\sigma \in \Sigma_F$
when the choice of $j$ is irrelevant. 

\ex{\label{thungy}
Set $d=3$: Suppose $F$ is a concrete loop configuration of form 
()(),  
that $D_e$ is a concrete diagram realising the trivial perm 
(derived from the trivial isotopy on $\R^2$, say) and 
that $D_{\sigma}$ realises the other perm
(derived from some other isotopy on $\R^2$, 
call it $k$). 
Then $D_e \simi D_{\sigma}$. 
Note that both are of duration 1 (i.e. $\tz = 1$) by construction.
An example of an isotopy $j_s$ realising the equivalence is
one which, at each time slice $\tz$, evolves linearly with $s$ between
the trivial isotopy on $\R^2$ and the homeomorphism $k_\tz$
from the family in $k$.
}




\newcommand{\PiF}{\Pi}

\begin{definition} \label{def:pif} \label{def:realises}
Let $\PiF_F \subseteq \Smin{d}[F,F]$ 
be the subset such that $D\in \PiF_F$ implies
$\conn(D)$ an isomorphism in $\CC{}_{\Pa}$ 
(cf. Proposition~\ref{pr:skel1}),
i.e. a permutation of $\Pt_1 F$. 
Thus $\pif_F \subset \PiF_F$. 
\\
A {\em complete} subset of $\PiF_F$ is one in which 
every $\sigma\in\Sigma_F$ is realised exactly once.
\label{def:xf}
\end{definition}

\newcommand{\sgi}{\eta}%

\begin{definition} 
Let $\Xi$ be a complete subset of $\Pi_F$. Define 
\begin{equation}
\sgi(\Xi):={1\over |\Sigma_F|}\sum_{X\in \Xi} X,
\label{eq:idemp}
\end{equation}
\label{def:ixf}
\end{definition}



\begin{lemma}
Let $A$ and $A'$ in $\Pi_F$ realise permutations $\sigma$ and $\sigma'$ 
respectively. Then $A\circ A'$ realises $\sigma'\sigma$.
\label{lem:concperm}
\end{lemma}

\noindent
{\em Proof: }
Follows directly from definition \ref{def:realises}. \Qed


\ochat{{
\[ \]
FIX ME: \hrule
\[ \]


Fix $F$. Let $X_F$ be a set of concrete diagrams realising the perms
$\Sigma_{F}$. 
For each $D \in \St{d}[F,F]$ 
define an element of $\Z \St{d}[F,F]$ by
\[
\bar{D} = \sum_{D_{\sigma} \in X_F} D \circ D_{\sigma} 
= D \circ e_F
\] 
where
\[
e_F = \sum_{D_{\sigma} \in X_F}  D_{\sigma} 
\]
(mutliplication by a Young symmetriser, as it were, 
although note that $e_F$ is not normalisable as an idempotent here). 
Note that all the summands pass to the same element under isotopy.

Example: [EXAMPLE]. 


\pr{ $\bar{D_i} \sim \bar{D_i'}$  implies 
$\bar{D_1} \circ \bar{D_2} \sim \bar{D_1'} \circ \bar{D_2'}$. }
{\em Proof:} 

[IDEA: compute products in concrete layer]
[BUT IS IT EVEN TRUE!!??...]


}}

\newcommand{\dd}{3} 
\section{Diagrams: classes of \CD s}

\noindent
In this section we define an equivalence relation on 
$\Smin{\dd}[F,F']$ 
(called {\it heterotopy}), whose equivalence classes will 
become the basis of a finite category
generalising  $\CC{}_T$.  
Note that, with $d>2$, isotopy classes are no longer big enough:


\pr{ {\rm \cite{AlvarezMartin07b}}
For $F \neq \emptyset$, 
$\Smin{\dd}[F,F]/\simi$ is a countably infinite set.
}

\ex{
There is an epimorphism from 
$\Smin{\dd}[(),()]/\simi$  
onto the set of knots.
}


For $d=2$ the category 
$\CC{d}_{\si}$
essentially coincides with $\CC{2}_{\ii}$. 
Both are infinite categories, but only because homs can contain
bubbles, as in Figure~\ref{D=}.
For $d>2$, $\CC{d}_{\si}$ is an infinite category, both by `knotting'
and by the formation of handles and bubbles. 
Our next objective is to give ways to eliminate these infinities
which will both connect with and generalise $\FF(\CC{d})$. 
We observe that the case for eliminating each bubble in favour of a
scalar, as in \TL\ (see Section~\ref{S:dichro}), 
is compelling. In Section~\ref{ss: het} we propose an analogous
treatment for `handles' in $d=3$,
which will also take care of knotting. 
Before that, we must prepare some machinery.



\subsection{Non-infinitesimal \CD\ transformations} \label{ss: ni}

\newcommand{\di}{\overline{D}}
\newcommand{\inter}{\mbox{int}}%

As already noted, each \CD\ $D \in \St{d}_t$ separates $E^d_t$ into 
connected components, 
with the point set $\Pt_0(D)$ of the diagram itself as the boundary. 
Let us call the connected components of  $E^d_t \setminus \Pt_0(D)$
the {\em alcoves} of $D$. 

Under suitable conditions the symmetric difference of two \CD s 
$D,D' \in \St{d}_t$ is again a \CD\ in $\St{d}_t$.
This is false if $\Pt_0(D)$ intersects 
more than one 
alcove of $D'$;
and trivially true if the point sets of $D$ and $D'$ do not intersect.
It can also be true if the point sets intersect in a disk or disks. 
There are some potential subtleties to this, but for
our purposes the following picture will be adequate.

For $T,T'$ sets,  define the {\em symmetric difference}
$$
T \swedge T' \; := \; T \cup T' \setminus T \cap T'
$$

\de{For $D,D' \in \St{d}_t$ we say they are $\swedge$-composable
(respectively  $\swedge^i$-composable) 
if $D \cap D'$ is a finite (or empty) union of disjoint disks
(respectively $i$ disjoint disks).
Noting that $D\swedge D'$ may be open due to the removal of these disks
we write $D \cwedge D'$ for the corresponding closure.}

\begin{figure}
\[
\includegraphics{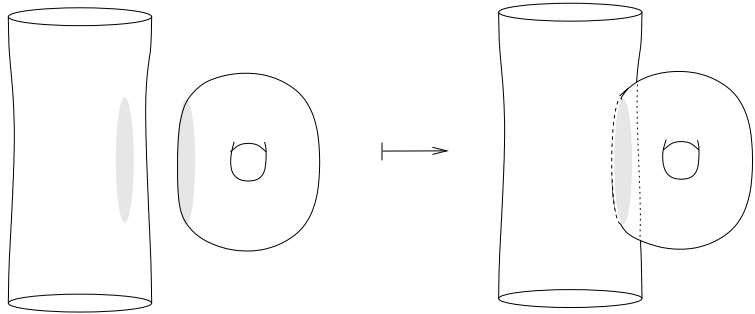}
\]
\caption{\label{wedge1} Example of $\cwedge$ product
(on the left the diagrams are drawn slightly separated, so that the
intersection is visible as a shaded disk in each).}
\end{figure}

\begin{figure}
\[
\includegraphics{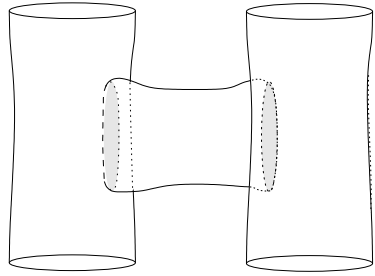}
\]
\caption{\label{wedge2} Example of $\cwedge$ product with $i=2$,
combining a diagram with two cylindrical components with a diagram
with one spherical component.}
\end{figure}

\begin{figure}
\[
\myfig{0}{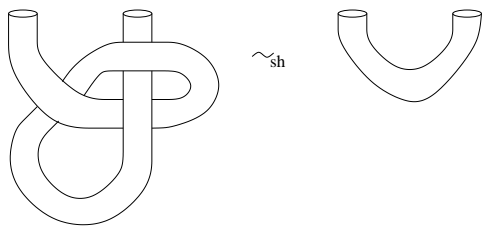}
\]
\caption{\label{stronghet1} 
NB, these \CD s are not strongly isotopic, or even isotopic.}
\end{figure}

\noindent
See Figure~\ref{wedge1} for an example with $i=1$, 
and Figure~\ref{wedge2} for $i=2$.


\lm{If $D,D'$  are $\swedge$-composable then  $D \cwedge D' \in \St{d}_t$.}
{\em Proof:} Since the intersection is made of
disks the interior of $D'$ is either entirely in the interior, or the
exterior, of $D$. If it is in the exterior, then   $D \cwedge D'$
separates  $E^d_t$ into an interior part which is the union of the
interiors and the open disks. 
If it is in the interior then the exterior of   $D \cwedge D'$
is the union of the exterior of $D$, the interior of $D'$, and the
open disks. \Qed

\de{For $B \in \St{d}_t$ define 
$\Dom^i B = \{ A \in \St{d}_t \; | \; A,B \mbox{ $\swedge^i$-composable}
  \}$
(the superscript $i$ may be omitted as above).
Define
\begin{eqnarray*}
\delta^i_B : \Dom^i B & \rightarrow & \St{d}_t \\
A & \mapsto & A \cwedge B
\end{eqnarray*}
(again the superscript may be omitted). 
}


\ex{ If  $T$ is a 2-torus, and intersects $D \in \Dom T \subset \St{3}_t$ 
in a single disk, then
$\delta_T = \delta^1_T$ 
has the effect of adding a handle to $D$.  
(See Figure~\ref{wedge1}.)

\label{re:bridge}
If $\asphere$ is a 2-sphere, then 
$\delta^2_{\asphere}$ has the effect of connecting (`bridging') two
  components if the two disks
($\di_1, \di_2$ say) are in distinct components (see Figure~\ref{wedge2});
or of introducing a handle if they are not.
}

\lm{ \label{exist ds}
{\rm (i)} If $c,d$ are adjacent components 
(bounding the same alcove)
in \CD\ $A$ then there is a sphere
$\asphere$ such that $\delta^2_{\asphere} A$ is a \CD\ differing from
  $A$ only in having a single component `composite' of $c$ and $d$. 
\\
{\rm (ii)} If $\asphere'$ is a second sphere similarly connecting $c$ and
    $d$ in $A$, and not intersecting $\asphere$, then 
$\delta^2_{\asphere'} \delta^2_{\asphere} A$ has a handle, and there is a
    torus $T$ such that 
\[
\delta^2_{\asphere'} \delta^2_{\asphere} A
\simsi
\delta^1_{T} \delta^2_{\asphere} A
\]
\Qed
}


\noindent
Define a relation $\rr_1$ on $S^3$ by $A \rr_1 B$ if 
there is a torus $T$ such that 
$B=\delta^1_T A$;
and 
$\rr$ as the transitive closure of $\rr_1$. 


If $B= \delta^1_T A$ then $B \cap T$ is a punctured torus (not a disk)
so that $B \not\in \Dom T$ in our definition. 
But $B\cwedge T =A$ so we can extend to allow the point-set operation
($\delta^{-1}_T$ by a mild abuse of notation) such that
$A=\delta^{-1}_T B$. 

We have
\[
g(\delta^1_T A ) = g(A)+1   \qquad   g(\delta^{-1}_T B ) = g(B)-1
\]

\pr{\label{pr isotopy d}
If $j$ is an isotopy and $A \in\Dom B$ then 
$j(\delta_B A)=\delta_{j(B)} j(A)$. Similarly 
$j\delta^{-1}_T B = \delta^{-1}_{j(T)} j(B)$.
\Qed
}

\subsection{\Heterotopy\ and \shet } \label{ss: het}

\ochat{
\futnote{
\de{ \label{rho def}
Define a relation $\rho$ on $\St3[F,F']$ by $X \rho X'$ if there exist
some set of tori such that 
\eql(ddd)
\delta_{T_n} \delta_{T_{n-1}} ... \delta_{T_1} X
= \delta_{T_n'} \delta_{T_{n-1}'} ... \delta_{T_1'} X'
\eq
and at every stage the intersection is a single disk.
Write $t=(T_1,T_2,...,T_n)$, $t'=(T'_1,T'_2,...,T'_n)$ and 
\[
X' = \rho_{t',t} X  \qquad \mbox{(or $X = \rho_{t,t'} X'$)} 
\]
as shorthand for the above identity. 
Thus $\rho_{t',t}$ encodes a specific `$\rho$-move' between two \CD s.
}
We write $X \in \Dom \rho_{t',t}$ if \CD\ $X$ may be related to
another \CD\ as in (\ref{ddd}). 
}
}

\newcommand{\kp}{\kappa}
\re{
There is no equivalent move to $ \delta^1_T$ 
for \TL\ diagrams, but a generalisation of
the move which replaces a closed loop in a \TL\ diagram with a scalar
factor would be to replace both bubbles (as in (\ref{bubblego})) and 
handles similarly.
Our next equivalence relation on $\St{\dd}$ will therefore be 
`handle replacement' --- meaning that if 
$B=\delta^1_T A$ in $\St{\dd}$ (or $\Ring\St{\dd}$) then
$B \equiv \kp A$ in a quotient of  $\Ring\St{\dd}$. 

First we construct a move which adds and removes equal numbers of
handles (thus generating an equivalence
bypassing the issue of scalars for now). 
}

\newcommand{\dt}[2]{\prod^{1}_{k=n} \delta^{#1_k}_{#2_k}}
\newcommand{\ssimh}{\approx_{\he}}
\newcommand{\ssimsh}{\approx_{\she}}

\begin{figure}
\[
\myfig{0}{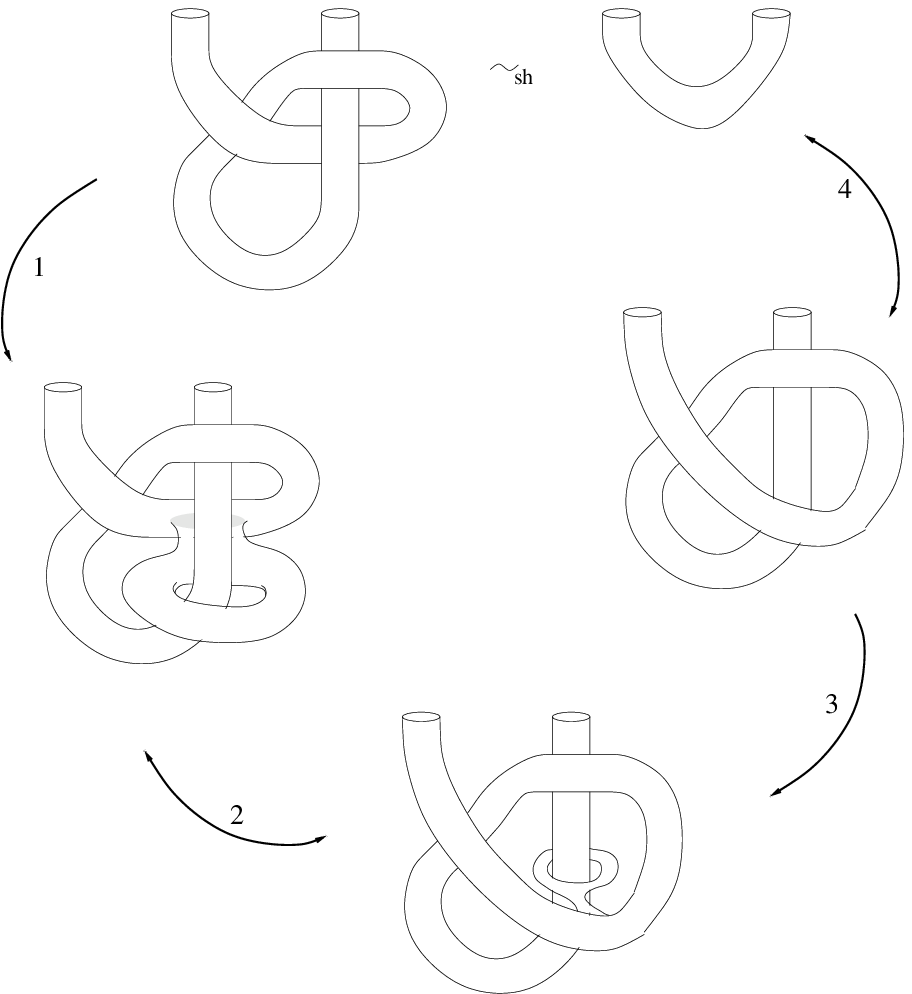}
\]
\caption{\label{stronghetoo1} 
The relation $A \simsh B$ 
in Figure~\ref{stronghet1}
is illustrated by the steps:
\newline
1.~$A \mapsto \delta^1_T A$; \newline
2.~$ \delta^1_T A  \mapsto i(\delta^1_T A) = \delta^1_{T'} i'(B)$;
\newline
3.~$  i'(B) \mapsto \delta^1_{T'} i'(B) =  i(\delta^1_T A)$;
\newline
4.~$ B \mapsto i'(B)$.
}
\end{figure}

\de{ \label{de: het}
A \het\  (\resp\ \shet) 
is a transformation on a \CD\ realised as  
a specific sequence of 
$\delta^{\pm 1}_T$-transformations and isotopies (\resp\ strong
isotopies). That is (noting Proposition~\ref{pr isotopy d}),
a \het\ is a transformation
\eql(hetseq)
A \mapsto B = i
    \dt{\epsilon}{T} A
\eq
for some $n$,
where $i$ is an (strong) isotopy, 
$t=(T_1,T_2,...,T_n)$ is a suitable set of tori, 
$\epsilon_k \in \{ \pm 1 \}$ and the order in the product matters.
\\
If such a  
transformation exists between
$A,B \in \Smin{\dd}$ we write  $A \simh B$ (\resp\ $A \simsh B$). 
This relation is an equivalence by construction. 
For $A \in \Smin{\dd}[F,F']$ 
we write $[A]_{\he}$ (\resp\ $[A]_{\she}$) for the equivalence class
of $A$ in $\Smin{\dd}[F'F']$, 
and call this class simply a {\em diagram}. 

The set of \het\ (\resp\ \shet) classes  
in $\Smin{\dd}[F,F']$ will be denoted
$\SSmin{\dd}[F,F']$ (\resp\  $\SSmins{\dd}[F,F']$).  

If such a  
transformation exists between $A,B \in \St{3}$ and 
 $\sum_k \epsilon_k = 0$ we write 
$A \ssimh B $  (respectively $A \ssimsh B $).
(So $A \simh B$ implies $A \ssimsh B $.)
}
Thus comparing (\ref{hetseq}) with Proposition~\ref{pr isotopy d}
and the definition of $\rr_1$:
\pr{
The equivalence relation in $\Smin{\dd}$ of {\em \het } 
(\resp\ \shet ) is  
the restriction of the RST closure of the 
relations $\rr_1$ and $\ii$ (\resp\ $\si$) on $\St{\dd}$ to
$\Smin{\dd}$.
A \het\ realising $A \simh B$ has $\sum_k \epsilon_k = 0$. 
\Qed
}
Examples: See Figure~\ref{stronghetoo1}.

\ochat{
\futnote{
NB --- THE FOLLOWING DISGAREES WITH THE OTHER PAPER!!!

\de{ \label{de: het}
A \het\  (\resp\ \shet) between two \CD s is a specific sequence of 
$\rho$-moves and isotopies (\resp\ strong isotopies).
If such a sequence exists between \CD s $A,B$ we write 
 $A \simh B$ (\resp\ $A \simsh B$). 
We write $[A]_{\he}$ (\resp\ $[A]_{\she}$) for the equivalence class of $A$. 

The equivalence relation in $\Smin{\dd}$ of {\em \het } 
(\resp\ \shet ) is thus the RST closure of the 
relations $\rho$ and isotopy (\resp\ strong isotopy).
}
Examples: For $\{i^0,i^1,...,i^m\}$ strong isotopies%
\footnote{We assume any (strong) heterotopy can be realised in finitely 
many moves.
}
\eql(real het)
B \; = \; 
i^0 \; \rho_{t^1,s^1} \; i^1 \rho_{t^2,s^2} ... i^{m-1} \; \rho_{t^m,s^m}
\; i^m \; A
\eq
realizes a \shet\ $A \simsh B$. 
\\
See Figure~\ref{stronghetoo1}.
}
}

\lm{ \label{st commute}
{\rm (i)} If sphere $\asphere$ and torus $T$ do not intersect 
and $A \in 
\Dom^2
{\asphere} \cap \Dom^1
{T}$ 
then  $\delta^1_{T} A \in  \Dom^2
{\asphere}$ and 
\[
\delta^2_{\asphere} \delta^1_{T} A = \delta^1_{T} \delta^2_{\asphere} A
\]
{\rm (ii)} Considering  (\ref{hetseq}) with $A \in \Smin{\dd}$,  
it is always possible to find a sphere $\asphere$ which bridges any two
adjacent components of $A $ but which intersects no torus in $t$.
Thus both $A,B$ in (\ref{hetseq}) lie in $\Dom^2 \asphere$. 
Further, if $i$ is strong then there is a neighbourhood of both
boundary (hyper)planes where it acts trivially, and hence 
$\delta^2_{\asphere}$ can be chosen to commute with $i$ also. 
}
{\em Proof:} (i) is trivial. (ii):  
None of the tori in $t$ touch the boundary (hyper)planes, so  
there is a neighbourhood of either where $\asphere$ can pass.
Since $A$ is minimal every component has a boundary component, and
we can bridge close to these (i.e. close to the boundary (hyper)plane). 
If the bridged components have boundary components on the same
boundary we are done; else by compactness we can chose the
inter-boundary part of $\asphere$ to be far away from any torus.
\Qed


\re{
Suppose tori $T,T'$ are isotopic and agree exactly except on a disk
$d=A\cap T$. Then $\delta^{-1}_{T'} \delta^1_T A$ is isotopic to $A$,
differing by the localised isotopy which replaces $d$ with $d'$ from
$T'$. 
Any isotopy $i$ can be realised by a sequence of such `local patch' 
moves, but it will be convenient and natural for us to keep isotopy as
a move itself. 
}

\section{Combinatorial characterisation of diagrams}

For \TL\ diagrams, 
which are isotopy classes of 
\CD s in $d=2$, we know that they can be placed in correspondence
with a subset of the set of pair partitions of their endpoints
--- a finite set. In $d=3$ we now have an analogous result. 

\begin{theorem}
Let $A,B \in \Smin{3}[F,F']$. Then $A \simsh B$ if and only if
$\conn(A)=\conn(B)$.
\label{prop:heteconn0}
\end{theorem}


\noindent
{\em Proof:} (Only if:)
Strong heterotopy is generated by `moves' none of which 
changes connectivity, hence if $A\simsh B$ then
$\conn(A)=\conn(B)$. 
\medskip

\noindent
(If:) We use a descending induction on the number of components of
$A$, with the maximum possible number $|A|=|F|+|F'|$ as base. 
In this case it is clear that $A\simsi B$.

Let P($k$) be the proposition that if 
$A,B \in \Smin3[F,F'](k)$ 
($k$ components), and $\conn(A)=\conn(B)$
then  $A\simsh B$. 
For the inductive step 
we require to show that P($k$) holds if P($k+1$) does. 

The strategy is to construct $A',B' \in \Smin3[F,F'](k+1)$ 
from $A,B$ (resp.) such that $\conn(A')=\conn(B')$, 
so that there is a heterotopy ($I$, say) between 
$A'$ and $ B'$, by the inductive assumption. 
Then from $I$ to construct a heterotopy between $A$ and $B$. 


\noindent
Step 1. Construction of $A',B'$

\noindent
In $A$ select a non-capped loop (labelled $l$, say) 
in $F$ that surrounds no other
non-capped loop (note that this is possible in all but the base case). 
Note that there is a (kind of singular limit of an isotopy) map 
which takes the collar to the
connected component at $l$ and pinches it, 
so yielding a cap at $l$ and a nearby patch of a now separate,
but {\em adjacent}
component (with no other components affected). 
We have:
\[
\includegraphics{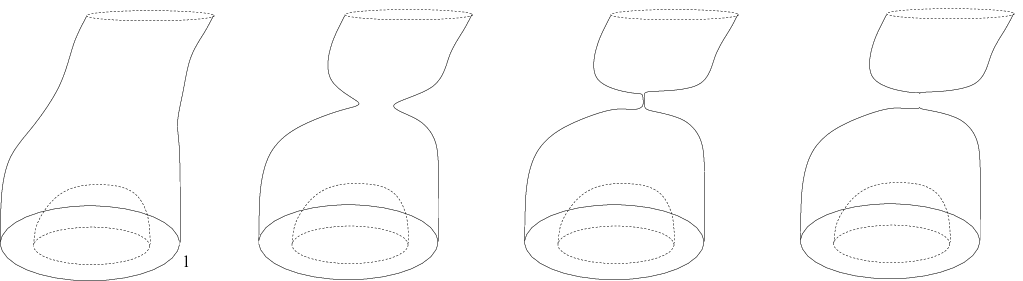}
\]
The construct $A'$ may be taken to be any such construct from $A$.
Note that $A' \in   \St3_0[F,F'](k+1)$.  

Note that in $B$ the loop $l$ is again non-capped, 
and again  surrounds no other non-capped loop. 
Accordingly construct $B'$ in the same way. 
Since $\conn(A)=\conn(B)$ and on each side all we have done is to move
the loop $l$ into a singleton part we have  $\conn(A')=\conn(B')$
as required. 

Let us use the label $c_l$ for the component of $B$ (or $A$) containing the
loop $l$; 
write $c_l'$ for the component of $B'$ (or $A'$) containing the
loop $l$; and $c_l''$ for the component of $B'$ (or $A'$) 
containing the other loops connected to $l$ in $c_l$. 


Note that there is a sphere $\asphere_A$ such that we may reconstruct $A$ from
$A'$ via
\eql(AA')
A = i_A \delta^2_{\asphere_A} A'
\eq
($i_A$ some isotopy), and similarly a  
sphere
$\asphere_B$ such that
\eql(BB')
B = i' \delta^2_{\asphere_B} B'
\eq 
That is, ${\asphere_B}$ meets $B'$ in a disk in component $c_l'$ and a
disk in component $c_l''$.


\medskip\noindent
Step 2. Construction of heterotopy between $A$ and $B$
\\
Since $\conn(A')=\conn(B')$,
by P($k+1$) we have  $A' \simsh B'$ (the inductive hypothesis). Let
\eql(BAAA)
B' \; = \; 
 i \dt{\epsilon}{T}  
\; A'
\eq
realize this \shet, as in (\ref{hetseq}).
(Note that $i$ is a {\em strong} isotopy here.) 


From (\ref{BB'}) and (\ref{BAAA}) we have 
\[
B = i' \delta^2_{\asphere_B} B' \; = \;  i' \delta^2_{\asphere_B} \;\; 
 i \dt{\epsilon}{T}  
\; A'
\]
The idea now is somehow 
to pass the $\delta^2_{\asphere_B}$ through to the right, 
and to end up with a heterotopy on $A$. 
To do this we need a couple of Lemmas.




\lm{ \label{lem 2301}
Suppose that a sphere $\asphere$ meets a \CD\ $X$ in a disk in some
 component $c$ and a disk in some component $d$.
Then for any other sphere $\asphere'$ with the same bridging property 
$\delta^2_{ \asphere} X \simsh \delta^2_{ \asphere'} X$. 
}
{\em Proof:}
We may assume that $s$ does not intersect all paths from $c$ to $d$ in 
the connected component of 
$E^{\dd}_t \setminus X$ 
containing its interior. Thus 
any such sphere $s'$ can  be isotopically deformed so as not to intersect 
 ${\asphere_{}}$, whereupon there are tori $T,U$ such that 
$\delta^1_{T}  \delta^2_{\asphere_{}} X 
\simsi \delta^2_{\asphere_{}'} \delta^2_{\asphere_{}} X
\simsi \delta^1_{U}  \delta^2_{\asphere_{}'} X $ 
($T$ meets $ \delta^2_{\asphere_{}} X $ in a disk consisting in a patch
 in $c$ in $X$; a strip  in $\asphere_{}$; and a patch in $d$ in
 $X$,
and $U$ is constructed similarly).
\Qed



\lm{ \label{lem d com}
Let $B= i \prod^1_{k=n} \delta^{\epsilon_k}_{T_k} A$ 
as in Definition~\ref{de: het}, 
and $c,d$ adjacent components in $A$
(and use the same labels for the corresponding components of $B$).
Then there exists a sphere $\asphere$ connecting $c$ to $d$
such that
\[
\delta^2_{\asphere} 
\dt{\epsilon}{T} A 
\; \simsi \;
\dt{\epsilon}{T} \delta^2_{\asphere} A
\]
}
{\em Proof:}
Noting Lemma~\ref{st commute}(i),
it is enough to show that there is such a sphere which does not intersect
any of the tori, but this was established in Lemma~\ref{st commute}(ii).
\Qed


\ochat{
OLD VERSION HOPING TO BE REMOVED!!:

\lm{ \label{lem d com}
Let $u,v$ be ordered sets of tori as in Definition~\ref{rho def}, 
$A \in \Dom\rho_{u,v}$,
and $c,d$ adjacent components in $A$
(and use the same labels for the corresponding components of $\rho_{u,v} A$).
Then there exist spheres $\asphere,\asphere'$ connecting $c$ to $d$
and sets of tori $u',v'$ such that
\[
\delta^2_{\asphere}\rho_{u,v} A \simsi \rho_{u',v'} \delta^2_{\asphere'} A
\]
}

\noindent
(Proof in Section~\ref{SSproof}.)

}
\newcommand{\prooflemdcom}{{
{\em Proof:}
Neither adding nor removing a handle necessitates a change in any
labelling of the connected components in a diagram, 
nor affects the property of adjacency of components; 
so we can also
consider neighbours $c,d$ in $\rho_{u,v} A$ corresponding to those in
$A$. Thus by Lemma~\ref{exist ds}(i) 
there are spheres $\asphere,\asphere'$ 
connecting $c$ to $d$ as appropriate. 
Note in particular that $\asphere'$ can be chosen so that 
$\asphere' \cap T_i = \emptyset$ for all $i$ (all tori in $v$). 
For that choice we have
\[
\delta_{\asphere'} \delta_{T_1} ... \delta_{T_n} A
=
 \delta_{T_1} ... \delta_{T_n} \delta_{\asphere'} A
\]
by Lemma~\ref{st commute}. 
We can similarly choose $\asphere$ to avoid the tori
$u=\{ S_1, ...,S_n \}$; and 
by definition
\eql(s-t)
\delta_{S_1} ... \delta_{S_n} \rho_{u,v} A
=
 \delta_{T_1} ... \delta_{T_n} A
\eq
 so similarly
\[
\delta_{\asphere} \delta_{S_1} ... \delta_{S_n}  \rho_{u,v} A
=
 \delta_{S_1} ... \delta_{S_n} \delta_{\asphere}  \rho_{u,v} A
\]

Note that $\asphere,\asphere'$ can also be chosen non-intersecting. 
Let us denote by $e$ the component resulting from the bridging of $c,d$ with
$\asphere$; and by $e'$ the result of the bridging of $c,d$ with
$\asphere'$. We have
\[
\delta_{\asphere'} \delta_{\asphere}  
\delta_{S_1} ... \delta_{S_n}  \rho_{u,v} A
=
\delta_{\asphere} \delta_{\asphere'}  
\delta_{S_1} ... \delta_{S_n}  \rho_{u,v} A
\stackrel{(\ref{s-t})}{=}
\delta_{\asphere} \delta_{\asphere'}  
 \delta_{T_1} ... \delta_{T_n} A
=
\delta_{\asphere}
 \delta_{T_1} ... \delta_{T_n}  \delta_{\asphere'}  A
\]
Here the $\delta_{\asphere}$ adds a redundant bridge to $e'$,
i.e. a handle. Evidently then, there is a torus $U$ (say) which would
add a strongly isotopic handle (Lemma~\ref{exist ds}(ii)):
\eql(2201)
\simsi \delta_U
\delta_{T_1} ... \delta_{T_n}  \delta_{\asphere'}  A
\eq
By a similar argument
\eql(2202)
\delta_{\asphere'} \delta_{\asphere}  
\delta_{S_1} ... \delta_{S_n}  \rho_{u,v} A
=\delta_{\asphere'}
\delta_{S_1} ... \delta_{S_n}  \delta_{\asphere}   \rho_{u,v} A
\simsi
\delta_{V}
\delta_{S_1} ... \delta_{S_n}  \delta_{\asphere}   \rho_{u,v} A
\eq
for some torus $V$. Combining these we have 
\[
i(\delta_U  \delta_{T_1} ... \delta_{T_n}  \delta_{\asphere'}  A)
= 
\delta_{V}
\delta_{S_1} ... \delta_{S_n}  \delta_{\asphere}   \rho_{u,v} A
\]
for some strong isotopy $i$,
i.e. that 
$ \delta_{\asphere}   \rho_{u,v} A$ and 
$i( \delta_{\asphere'}  A)$ are related by a $\rho$-move
as in Definition~\ref{rho def}. 
After applying Proposition~\ref{pr isotopy d} 
a couple of times we can write this in the form:
\[
 \delta_{\asphere}   \rho_{u,v} A
\simsi
\rho_{\tilde{Vu},Uv}
 \delta_{\asphere'}  A
\]
(using an abreviated notation for the new $\rho$-move,
the details of which are not needed). \Qed
}}


The sphere $\asphere_B$ in (\ref{BB'}) is not 
necessarily an entirely free choice, but by Lemma~\ref{lem 2301},
replacing  (\ref{BB'}) with  
$B \simsh  i' \delta^2_{\asphere_{}} B'  $
we can choose $\asphere$ as in Lemma~\ref{lem d com}. 
In particular we can choose $s$ close to a boundary hyperplane,
so that it commutes with strong isotopy $i$ also. 
We have
\[
B \simsh
 i' \delta^2_{\asphere_{}}  \; i \; 
\dt{\epsilon}{T}
\;  \; A'
\simsi
 i'  \; i \; 
\dt{\epsilon}{T}
\; \delta^2_{\asphere_{}} \;  \; A'
\]
That is $B \simsh  
\; \delta^2_{\asphere} \; A'$. 
But using Lemma~\ref{lem 2301} again 
$ \delta^2_{\asphere} \; A' \simsh \delta_{\asphere_A} A' 
\simsi A$.
We have established that P($k+1$) implies P($k$) as required. 
\Qed


\section{The \shet\ diagram category}\label{s:sh}

\newcommand{\dq}{q}
\newcommand{\circsh}{\circ_{\she}}

\pr{ 
Fix $F,F'$. 
For each $A \in \St{\dd}[F,F']$ the set 
$\{ D \in \Smin{\dd} \; | \; D \rr \Rem(A) \}$
is non-empty and 
lies within a single \shet\ class 
(and hence also \het\ class in $\Smin{\dd}[F,F']$).
Call this class $\Down_{\she} (A)$ (\resp\ $\Down_{} (A)$),
then we have a surjective map
\[
\Down_{\she} : \St{3} \rightarrow \St{3}_{\she} .
\] 
}
{\em Proof:} 
Recall that the \CD\ $\Rem(A)$ has no bubbles. 
Note also that every \CD\ with a handle has 
a neighbourhood (of some such handle) isotopic to that illustrated in
Figure~\ref{handle1}, and hence has the
relation $\rr_1$ with a \CD\ with one fewer handle.
Thus there is a $D\in\Smin{\dd}$ satisfying $D \rr \Rem(A)$.  
The inclusion follows from the various definitions involved 
(in particular definition~\ref{de: het} (of $\simh$ and $\simsh$)). 
The final part is clear.
\Qed



\begin{definition}
Fix $\K$ a field and $\kp,q\in\K$.
Define
$\mu = \mu_{\kp q}$ and $\mu^{\she}_{}$ by
\[
\begin{array}{ccccccc}
\mu_{\kp \dq}: \St{\dd}[F,F'] &\longrightarrow&  \K \SSmin{\dd}[F,F']
&\qquad&
\mu^{\she}_{}: \St{\dd}[F,F'] &\longrightarrow&  \K \SSmins{\dd}[F,F']
\\
\hfill A  & \mapsto &   \kp^{g(A)} \dq^{b(A)} {\Down}(A)
& \qquad &
\hfill A  & \mapsto &   \kp^{g(A)} \dq^{b(A)} {\Down}_{\she}(A) 
\end{array}
\]
and extend the domain of $\mu^{\she}$ (resp. $\mu$)
linearly to $\K\St{\dd}[F,F']$.
\label{def:mupq}
\end{definition}


\pr{
If $A,A' \in \Smin{3}$ then $A \simsh A'$ if and only if 
$\mu^{\she}(A) = \mu^{\she}(A')$.
If  $A,A' \in \St{3}$ then $A \ssimsh A'$ implies 
$\mu^{\she}(A) = \mu^{\she}(A')$.
}
{\em Proof:} 
If  $A,A' \in \Smin{3}$ then $\Down_{\she}(A)=[A]_{\she}$ and 
 $\Down_{\she}(A')=[A']_{\she}$, 
and   $b(A)=b(A')=g(A)=g(A')=0$.
\\
For the second case, 
note that $A \ssimsh A'$ implies $b(A)=b(A')$,
$g(A)=g(A')$. 
\Qed

Let $A,A' \in \Smin{\dd}[F'',F]$  and $B,B' \in \Smin{\dd}[F,F']$.
If $A \simsh A'$ and $B \simsh B'$ then 
$A\circ B \ssimsh A' \circ B'$
(stack the transformations just as the \CD s are stacked). 
It follows that 
\pr{
$\mu^{\she}(A\circ B)$ depends on $A$ and $B$ only through their
$\she$-classes.
}  
Hence  there is a well-defined composition  
$\circsh : \K \SSmins{\dd}[-,F] \times \K \SSmins{\dd}[F,-]
\rightarrow \K \SSmins{\dd}[-,-]$ 
given by
\eql(circsh)
[A]_{\she} \circsh [B]_{\she} \; := \; \mu^{\she}(A\circ B)
\eq
{\theor{
The triple 
\[
\K \CC{}_{\she} =\K \CC{}_{\she} (\kp,q)
= ( \Sto{2} ,\K \St{\dd}_{\she}[-,-] ,  \circsh)
\]
is a category.
With $\kp=1$ it is isomorphic to a subcategory of $\K\CC{}_{\Pa(q)}$.
For general $\kp$ it is a deformation of this subcategory. 
}}
{\em Proof:} The equivalence relation on  $\K \St{\dd}[-,-]$ given by 
$\mu^{\she}$ agrees with $\simsh$ on $\Smin{\dd}$. 
The well-defined composition (\ref{circsh}) thus extends to a congruence.
The new category is a quotient of $\K \CC{\dd}$ by this congruence.
\Qed


\pr{\label{pr:isoisot}
The hom set $\K \St{\dd}_{\she}[F,F']$ contains isomorphisms if and
only if $F \simi F'$. 
If $j_s$ is an isotopy and $F'=j_1(F)$ then 
$[D^{j_s}_F]_{\she}$ is an isomorphism in $\K\St{\dd}_{\she}[F,F']$. 
}
In consequence a skeleton for $\K \CC{}_{\she}$ has object set 
in bijection with the set
of rooted trees. 
The category $\K\CC{}_{\she}$
has an intriguing representation theory, that we shall
return to shortly. 


\section{A \het\ category}
\subsection{Composition of diagrams in $\K \SSmin{\dd}$}

In this section we introduce a composition $.$ 
making the triple 
$
\CC{}_{\he}=(\Sto{2},\K \SSmin{\dd}[-,-],.)
$ 
a category. 


We will show shortly (in Theorem~\ref{th:basicth}) the following:
\\
Let $A,A' \in \Smin{\dd}[F'',F]$  and $B,B' \in \Smin{\dd}[F,F']$.
Let $X_F$ be any complete subset of $\Pi_F$. 
If $A \simh A'$ and $B \simh B'$ then 
\begin{eqnarray*}
\mu_{\kp\dq}(A\circ \sgi(X_F)\circ B)
 = \mu_{\kp\dq}(A'\circ \sgi(X_F)\circ B').
\end{eqnarray*}
In other words $\mu_{\kp\dq}(A\circ \sgi(X_F)\circ B)$ depends on $A,B$
only through their $\he$-classes. 
This makes the following construction well defined.

\newcommand{\hec}[1]{[#1 ]_{\he}}

\begin{definition} \label{def:compodef}
Let $\hec{ A} \in \SSmin{\dd}[F'',F]$ and 
$ \hec{ B} \in \SSmin{\dd}[F,F']$, 
and 
let $A, B$ be {\em any} minimal concrete diagrams in $\hec{ A}$ and $\hec{ B}$ 
respectively. 
Let $X_F$ be {\em any} complete subset of $\Pi_F$. 
Then
\begin{eqnarray*}
\hec{ A} \cdot\hec{ B}
  = \mu_{\kp\dq}\left(A\circ \sgi(X_F)\circ B \right)
\in \K \SSmin{\dd}[F'',F'] .
\end{eqnarray*}
\end{definition}

In fact we will show (equation~(\ref{eq:aixfb})) that this composition
does not depend on the choice of $X_F$ either. 
Thus $\hec{ A}\cdot\hec{ B}$
is natural and 
well defined in that, for given $p$ and $q$ in field $\K$, it depends 
only on $\hec{ A}$ and $\hec{ B}$.


We next prove the well-definedness theorem; and then turn to 
study the properties of this composition. 

\subsection{Well-definedness Theorem}

\begin{proposition} \label{prop:hassh}
Let $A,B$ in $\Smin{d}[F,F']$.  
Then $A \simh B$ if and only if 
there are $L$ in $\Pi_F$ and $R$ in $\Pi_{F'}$ such that
\begin{eqnarray*}
A\simsh L\circ B\circ R.
\end{eqnarray*}
The same holds with $\he$ replaced by $\ii$ and $\she$ by $\si$.
\end{proposition}
\oproof{Def~\ref{def:unitf}}{ 
Noting Definition~\ref{def:unitf}
we have that 
$A$ is strongly heterotopic (in fact, strongly isotopic) to 
$I_F\circ A\circ I_{F'}$. 
Apply to $A$ the heterotopy that takes it to $B$
and simultaneously change the $I_F,I_{F'}$ 
by isotopies in a small neighbourhood 
of $A$ so that the image of $I_F\circ A\circ I_{F'}$ under these operations
is a concrete diagram at all times. Under those isotopies, the $I_F$ and 
 $I_{F'}$ become the required $L$ and $R$ respectively. The result is a
strong heterotopy from $I_F\circ A\circ I_{F'}$ to $L\circ B\circ R$ and the
proposition follows.
}

\begin{lemma}
Let $A$ be in $\St{d}[F,F']$ (note: not necessarily minimal). For any $L$ 
in $\Pi_F$ 
and $R$ in $\Pi_{F'}$, we have $\mu_{pq}(A)=\mu_{pq}(L\circ A\circ R)$.
\label{lem:muamular}
\end{lemma}
\oproof{Prop~\ref{prop:hassh}}{
Let $C \in \Down(A)$. Then $L\circ C\circ R \in 
\Down( L\circ A\circ R )$. 
By Proposition~\ref{prop:hassh} 
we have $C\simh L\circ C\circ R$ and 
hence  
$\mu_{pq}(C)=\mu_{pq}(L\circ C\circ R)$. 
By the definition of $\mu_{pq}$,
\begin{eqnarray*}
\mu_{pq}(A)&=&p^{g(A)}q^{b(A)}\mu_{pq}(C),\\
\mu_{pq}(L\circ A\circ R)&=&
p^{g(L\circ A\circ R)}q^{b(L\circ A\circ R)}\mu_{pq}(L\circ C\circ R).
\end{eqnarray*}
But clearly $A$ and $L\circ A\circ R$ have the same number 
of handles and bubbles.
}


\begin{proposition} \label{prop:concsh}
Let $A_{i},A_{i}' \in \Smin{d}[F^i,F^{i+1}]$ for $i=1,2,...,n$.
If $A_i \simsh A_{i}'$ then 
\begin{equation}
\label{eq:multicirc}
\mu_{pq}(A_1\circ\ldots\circ A_n)=\mu_{pq}(A'_1\circ\ldots\circ A'_n) .
\end{equation}
\end{proposition}
\oproof{Claim~\ref{claim xcon}}{
Let $C\in \Down(A_1\circ A_2)$ and $C'\in \Down(A_1'\circ A_2')$. 
By Proposition~\ref{claim xcon}  
$\conn(A_1\circ A_2)=\conn(A_1'\circ A_2')$, 
therefore $\conn(C)=\conn(C')$ and by 
Theorem~\ref{prop:heteconn0}
we have $C\simsh C'$. Now,
\begin{eqnarray*}
\mu_{pq}(A_1\circ A_2)&=&p^{g(A_1\circ A_2)}q^{b(A_1\circ A_2)}\mu_{pq}(C),\\
\mu_{pq}(A_1'\circ A_2')&=&
p^{g(A_1'\circ A_2')}q^{b(A_1'\circ A_2')}\mu_{pq}(C').
\end{eqnarray*}
But $\mu_{pq}(C)=\mu_{pq}(C')$ and the number of bubbles and handles
is the same in $A_1\circ A_2$ and in $A_1'\circ A_2'$. 
Now iterate.
}


\pr{
If $X_F$ and $X'_F$ are two complete subsets of 
$\Pi_F$, and $\sgi(X_F)$, $\sgi(X'_F)$ the corresponding idempotents (see 
definitions \ref{def:xf} and \ref{def:ixf}) then for any 
$A \in \Smin{d}[F'',F]$  and $B \in \Smin{d}[F,F']$,
\begin{equation}
\mu_{pq}(A\circ \sgi(X_F)\circ B)
 = \mu_{pq}(A\circ \sgi(X'_F)\circ B)
\label{eq:aixfb}
\end{equation}
}
\proof{
For every term in $\sgi(X_F)$ there is  a term in $\sgi(X'_F)$ that
realises the same permutation, hence has the same connectivity.
By Theorem~\ref{prop:heteconn0} 
those two terms are strongly heterotopic minimal diagrams.
Now apply equation~(\ref{eq:multicirc}) in case $n=3$, and the 
linearity of $\mu_{pq}$.
}


\begin{theorem} \label{th:basicth}
Let $A,A' \in \Smin{d}[F'',F]$  and $B,B' \in \Smin{d}[F,F']$.
Let $X_F$ be any complete subset of $\Pi_F$. 
If $A \simh A'$ and $B \simh B'$ then 
\begin{eqnarray*}
\mu_{pq}(A\circ \sgi(X_F)\circ B)
 = \mu_{pq}(A'\circ \sgi(X_F)\circ B').
\end{eqnarray*}
\end{theorem}
\proof{
By proposition \ref{prop:hassh} there are $L_1,R_1,L_2,R_2$ in $\Pi_F$ 
such that
\begin{eqnarray*}
A &\sim_{sh}& L_1\circ A'\circ R_1,\\
B &\sim_{sh}& L_2\circ B'\circ R_2.
\end{eqnarray*}
Therefore, by equation (\ref{eq:multicirc}),
\begin{eqnarray*}
\mu_{pq}(A\circ \sgi(X_F)\circ B)
 = \mu_{pq}(L_1\circ A'\circ R_1\circ \sgi(X_F)
\circ L_2\circ B'\circ R_2).
\end{eqnarray*}
By lemma \ref{lem:muamular} (which allows $L_1$ and $R_2$ to be eliminated),
\begin{eqnarray*}
\mu_{pq}(A\circ \sgi(X_F)\circ B)
 = \mu_{pq}(A'\circ R_1\circ \sgi(X_F) \circ L_2\circ B').
\end{eqnarray*}
But $R_1\circ \sgi(X_F)\circ L_2$ is strongly heterotopic 
(in the obvious sense)
to $\sgi(X_F)$, hence
applying equation (\ref{eq:multicirc}) again we 
are done. 
}


\subsection{Properties of the composition $\hec{ A}\cdot\hec{ B}$}

\begin{theorem}
\label{th:het cat}
The triple
\[
\CC{}_{\he}=(\Sto{2},\K \SSmin{\dd}[-,-],.)
\]
is a category.
\end{theorem}
{\em Proof:}
We need to check (i) for identity elements, and (ii) for associativity.
\\
(i)
For any $\hec{ A} \in \SSmin{\dd}[F,F]$, 
and any $X_F$ a complete subset of $\Pi_F$, we have
\newcommand{\hone}[1]{[ I_{#1} ]_{\he}}
\begin{eqnarray*}
\hec{ A} \cdot \hone{F}
= \mu_{pq}\left(A\circ \sgi(X_F)\circ I_F \right)
= \mu_{pq}\left(A\circ \sgi(X_F)\right)=\mu_{pq}(A)
=\hec{ A}.
\end{eqnarray*}
Similarly, $\hone{F} \cdot\hec{ A}=\hec{ A}$. 
\\
(ii) It remains to prove associativity:

\begin{lemma} \label{th:assoccomp}
For any $\hec{ A}, \hec{ B}, \hec{ C}\in \SSmin{3}[-,-]$
with $A,B,C$ composable,
\begin{eqnarray*}
(\hec{ A}\cdot\hec{ B})\cdot\hec{ C}=
\hec{ A}\cdot(\hec{ B}\cdot\hec{ C}).
\end{eqnarray*}
\end{lemma}
\proof{
Let $A,B,C $ be composable in $ \Smin{3}$, and $X_F,X_{F'}$ complete.
Let $D_X \in \Down(A\circ X\circ B)$ for $X \in X_F$,
$G_{X'} \in \Down(B\circ X' \circ C)$ for $X' \in X_{F'}$, and 
$E_{XX'} \in \Down(D_X\circ X'\circ C)$ and 
$H_{XX'} \in \Down(A\circ X\circ G_{X'})$.
Then
\begin{eqnarray*}
[A]_{\he} \cdot [B]_{\he} 
&=&{1\over |\Sigma_F|}\sum_{X\in X_F}\mu_{pq}
(A\circ X\circ B)\\ &=&{1\over |\Sigma_F|}\sum_{X\in X_F}
p^{g(A\circ X\circ B)}q^{b(A\circ X\circ B)} [D_X]_{\he} 
\end{eqnarray*}
so
\begin{eqnarray*}
([A]_{\he} \cdot [B]_{\he} ) \cdot [C]_{\he}
=
{1\over |\Sigma_F|^2}\sum_{X\in X_F}
p^{g(A\circ X\circ B)}q^{b(A\circ X\circ B)}
\sum_{X'\in X_{F'}}p^{g(D_X\circ X'\circ C)}q^{b(D_X\circ X'\circ C)}
[{E}_{XX'}]_{\he} ,
\end{eqnarray*}
\begin{eqnarray*}
[A]_{\he} \cdot ( [B]_{\he}  \cdot [C]_{\he} )
=
{1\over |\Sigma_F|^2}\sum_{X\in X_F}
p^{g(A\circ X\circ G_{X'})}q^{b(A\circ X\circ G_{X'})}
\sum_{X'\in X_{F'}}p^{g(B\circ X'\circ C)}q^{b(B\circ X'\circ C)}
[{H}_{XX'}]_{\he},
\end{eqnarray*}
Note that $H_{XX'},E_{XX'}   \in
\Down( A\circ X\circ B\circ X'\circ C)$, 
and hence   
$[{ E}_{XX'}]_{\he} = [{ H}_{XX'}]_{\he}$.
It thus suffices to show that for each $X,X'$
in the double sums above:
\begin{eqnarray*}
g(A\circ X\circ B)+g(D_X\circ X'\circ C)&=&
g(A\circ X\circ G_{X'})+g(B\circ X'\circ C),\\
b(A\circ X\circ B)+b(D_X\circ X'\circ C)&=&
b(A\circ X\circ G_{X'})+b(B\circ X'\circ C).
\end{eqnarray*}
This follows from equations (\ref{eq:samebub}) and (\ref{eq:sameg}) 
(see Section~\ref{sec:genius}) applied 
to $A\circ X$, $B$ and $X'\circ C$.
}

\newcommand{\appendixCC}{{
\section{Lemmas for proof of associativity: genus counting}
\label{sec:genius}

If $a$ is a component of $A \in \St{\dd}$, 
then $\chi(a)$, $g(a)$ and $h(a)$ are the 
Euler number, genus, and number of holes in $a$. Then 
\begin{equation} \label{eq:chigh}
\chi(a)=2-2g(a)-h(a) .
\end{equation}


If $a_1,\cdots,a_n$ are the components of $A$, then
\begin{eqnarray}
\chi(A) &=& \sum_{i=1}^n \chi(a_i),
\label{eq:chiofa} \\
g(A)    &=& \sum_{i=1}^n g(a_i).
\label{eq:gofa}
\end{eqnarray}

It is clear that, if 
$A\in \Smin{3}[F,F']$   
then $g(A)=0$ and $\chi(A)=2|A|-(|F|+|F'|)$. 


\begin{lemma} \label{prop:gacircb}   
For $A \in \St{3}[F',F]$, $B\in\St{3}[F,F'']$ 
\begin{equation} \label{eq:lemchi}
\chi(A\circ B)=\chi(A) + \chi(B) .
\end{equation}
\begin{equation} \label{eq:gab}
g(A\circ B)=g(A)+g(B)+|A\circ B|-|A|-|B|+|F| .
\end{equation}
\end{lemma}
\proof{
(i) By \cite{Milnor63} 
$\chi(A)$ is the sum of the Morse indices of all the critical
points in the components of $A$, and similarly for $\chi(B)$ and 
$\chi(A\circ B)$. But the critical points of $A\circ B$ are those of $A$
plus those of $B$, with the same Morse indices. 
\\
(ii) From equations (\ref{eq:chiofa}) and (\ref{eq:gofa}), 
\begin{eqnarray*}
\chi(A) &=& 2|A|-2g(A)-(|F|+|F'|),\\
\chi(B) &=& 2|B|-2g(B)-(|F|+|F''|),\\
\chi(A\circ B) &=& 2|A\circ B|-2g(A\circ B)-(|F'|+|F''|) .
\end{eqnarray*}
Then the Lemma 
follows                   
immediately.         
}

Note also the following generalisation with $C\in\St{\dd}[F'',F''']$:
\begin{equation}
g(A\circ B\circ C)=g(A)+g(B)+g(C)+|A\circ B\circ C|-|A|-|B|-|C|+
|F|+|F''|   
\label{eq:gabc}
\end{equation}
for which we make use of associativity of concatenation.


\ochat{{
We mentioned above the operation on \TL\ diagrams where the concatenation
of two minimal diagrams is made minimal by removing closed loops. We now 
present an analogous construction in $S^3[F,F]$.

Note that all $B$ in $min(A)$ are {\it minimal}. We can think of the elements
of $min(A)$ as obtained from $A$ by removing all ``handles'' first, and then
removing all components which no boundaries (which we shall call ``bubbles''.)
Note also that there are in general many different ways to remove handles, 
a question whose implications will be discussed in the next question.

\bigskip
{\bf ----- Illustrate handles and bubbles with a figure -----}
\bigskip

Clearly, all diagrams in $min(A)$ have the same number of 
components.

\begin{definition}
For $A$ in $S^3[F,F]$, the {\it number of bubbles} of $A$, written $b(A)$,
is the number of components of $A$ without boundaries.
\label{def:bubbles}
\end{definition}

In particular, if $A$ is minimal, then $b(A)=0$. More generally, we have the
following Lemma:

}}


\begin{definition}
For $A$ in $\St{d}[F,F']$, 
define $min(A)$ as the set of $B\in \Smin{d}[F,F']$
having the same connectivity as $A$:
\begin{equation}
min(A)=\left\{ B\in \Smin{d} \>:\> \conn(A)=\conn(B) \right\}. 
\label{eq:mina}
\end{equation}
\label{def:minofa}
\end{definition}
\begin{proposition} 
\label{prop:bubabc} 
Let $A,B,C \in \Smin{3}$ be composable. 
Let $D\in min(A\circ B)$ and $G\in min(B\circ C)$. Then
\begin{equation} \label{eq:bacircb}
b(A\circ B)=|A\circ B|-|D| ,
\end{equation}
\begin{equation} \label{eq:samebub}
|A\circ B|-|D|+|D\circ C|=|B\circ C|-|G|+|A\circ G|,
\end{equation}
\begin{equation} \label{eq:sameg}
g(A\circ B)+g(D\circ C)=g(B\circ C)+g(A\circ G) .
\end{equation}
\end{proposition}
\proof{
(i) $|D|=|p(A \circ B)|$ by definition.
\\
(ii) Let $E\in min(D\circ C)$. Then $E$ is minimal and has the 
same connectivity of $D\circ C$. But the connectivity of $D\circ C$ is that
of $A\circ B\circ C$, so that $E\in min(A\circ B\circ C)$. Now, if
$H\in min(A\circ G)$, the same argument shows that 
$H\in min(A\circ B\circ C)$. Hence we can take $E=H$ without loss of 
generality. Now, the total number of bubbles in $A\circ B\circ C$ can
be counted as follows: those created in $A\circ B$, plus the new ones 
created in $(A\circ B)\circ C$. The latter are also those created in
the concatenation $D\circ C$, so that
\begin{eqnarray*}
b(A\circ B\circ C)=|A\circ B|-|D| + |D\circ C|-|E|.
\end{eqnarray*}

This number can also be counted by first concatenating $B\circ C$, then 
$A\circ G$, and finally minimising to $E$. Then
\begin{eqnarray*}
b(A\circ B\circ C)=|B\circ C|-|G| + |A\circ G|-|E|,
\end{eqnarray*}
which proves the Proposition part (ii).

\noindent
(iii) By Lemma \ref{prop:gacircb}
\begin{eqnarray*}
g(A\circ B)+g(D\circ C)=|A\circ B|-|A|-|B|+|D\circ C|-|D|-|C|+
|F|+|F''| 
\end{eqnarray*}
\begin{eqnarray*}
g(B\circ C)+g(A\circ G)=|B\circ C|-|B|-|C|+|A\circ G|-|A|-|G|+
|F|+|F''|.
\end{eqnarray*}
These expressions are equal by Proposition~\ref{prop:bubabc}(ii).
}

This equality may also be understood as two different ways to count
the number of handles in $A\circ B\circ C$. 

}}
Having constructed our categories, we now provide the basic tools
for practical computation within them.

\section{Practical enumeration of diagrams}


Embedded manifolds are not easy to manipulate combinatorially in
general. However, 
by Theorem~\ref{prop:heteconn0} we have an injective map 
$p: \St{3}_{\she}[F,F'] \rightarrow \Pa(F \dotcup F')$
for any pair $F,F'$.
The combinatorics of set partitions are quite well understood
--- the elements of $\Pa(F \dotcup F')$ are enumerated (for any
given enumeration of $F \dotcup F'$) in, for example,
\cite{Martin91}. 
Thus to enumerate $\St{3}_{\she}[F,F']$
it is sufficient to describe the subset $p(\St{3}_{\she}[F,F'])$
of  $\Pa(F \dotcup F')$. 
We do this next. 
(We then demonstrate the utility of the method by computing some
explicit multiplication tables in the heterotopy category.
We postpone detailed representation theory to a separate paper.)

In order to give an explicit 
combinatorial characterisation of diagrams, 
that is, a condition for a partition in $\Pa(F \dotcup F')$
to be in the image of the map $p$ above, 
 it will be helpful to recall some graph theory.


\subsection{Graph basics, tree graphs and $\Sto{2}$}

\de{
A (directed) graph $G$ is two sets, $V_G,E_G$ 
(the set of vertices and the set of edges, respectively), 
together with two functions 
$i:E_G \rightarrow V_G$ and $f:E_G \rightarrow V_G$.
\\
For each total order $\omega:V_G \rightarrow \N$ there is a matrix 
$\Omega_{\omega}(G)$ whose $j,k$-th entry is the number of 
edges $e$ such that $\omega(i(e))=j$ and  $\omega(f(e))=k$. 
\\
An undirected graph is a graph $G$ in which for every $e_1 \in E_G$ with 
$i(e_1)=v_1$ and $f(e_1)=v_2$ there is an  $e_1'$ with 
$i(e_1')=v_2$ and $f(e_1')=v_1$.
}

\de{
Graphs $G,G'$ are said to be {\em isomorphic} 
if for any $w$ a total order of $V_G$ there is a total order $w'$
of $V_{G'}$ such that $\Omega_{w}(G)= \Omega_{w'}(G')$.  
}

\de{ \label{col graph}
Let $G$ be an undirected graph and $C$ a set. 
An edge colouring of $G$ by $C$ is a map from the edge set of $G$ to
$C$. Write $G^C$ for the set of all edge colourings of $G$ by $C$. 
}
Note that any function $f$ defines a partition of its domain by 
$x \sim y$ if $f(x)=f(y)$. Thus we have a map from  $G^C$ 
to partitions of $E_G$. This map is surjective (each perm of the set
$C$ defines the same partition). 

\de{
A rooted graph is a pair $(G,v)$ consisting of a graph $G$ and an
element of $V_G$ (called the root). 
\\
Two rooted graphs  $(G,v)$, $(G',v')$
are isomorphic if they are isomorphic as graphs via an isomorphism 
in which the positions of the roots agree in the respective orders
(i.e. $w(v)=w'(v')$). 
}

Write $[G,v]$ for the isomorphism class of rooted graph $(G,v)$. 


\de{\label{de:rooted}
A rooted tree is a tree graph with a single distinguished vertex 
(others unlabeled),
that is, a class $[G,v]$ where $G$ is a tree graph.
\\
Write $\tree$ for the set of rooted trees and $\tree_n$  
for the set of rooted trees with $n$ vertices.
}

The association of $[G,v]$ to $(G,v)$ 
associates a rooted tree to each rooted tree graph 
by `forgetting' the labels on all the vertices except the root.

For $V$ a set let $\Graph(V)$ denote the set of graphs with vertex set
$V$. Let $\Graph_u(V)$ denote the subset of undirected graphs. 
Write $\Graph_u$ for the class of all finite undirected graphs;
and $\Graph_{ru}$ for the class of all finite undirected rooted graphs. 


Define 
\[
\gtree : \Sto{2} \rightarrow \Graph_{ru}
\]
as follows: 
(1) The vertex set of  $\gtree(F)$ 
is the set of connected components of 
$\Re^2 \setminus F$ (`regions'); 
(2) The root is the vertex associated to the unbounded region;
(3) There is an undirected edge between $v_1$ and $v_2$ 
in $\gtree(F)$ if there is 
a component in $F$ which is a boundary between the corresponding two
regions. 

It will be evident that this graph $\gtree(F)$ is a tree graph. 
(It is not, however,  
a rooted tree, since its vertices are labeled.)
\pr{\label{pr:bcrt}
Boundary configurations $F,F' \in \Sto{2}$ 
are isotopic if and only if their tree graphs pass to the same rooted tree,
i.e. $[\gtree(F)] = [\gtree(F')]$. \Qed
}
The passage from boundary configuration to rooted tree is exemplified in Figure 16.

\begin{figure}
\includegraphics{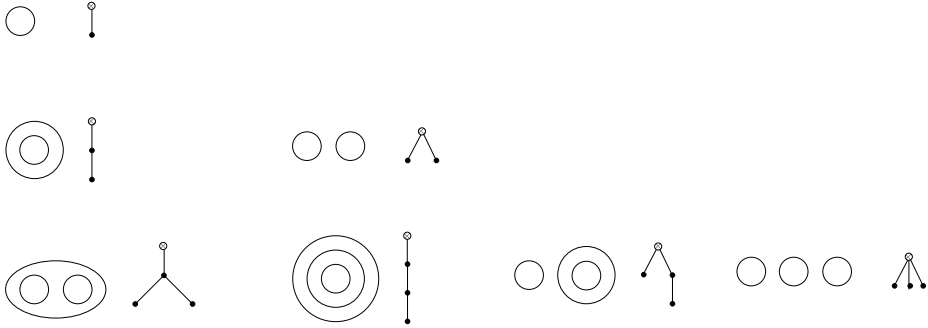}
\caption{\label{trees} Some boundary configurations in $\Sto{2}$  and associated trees. }
\end{figure}
\subsection{Enumerating diagrams in $\St{3}_{\she}$ and  $\St{3}_{\he}$}

Suppose that $S$ is a totally ordered set. The total order on $S$
induces a total order on the parts of any partition of $S$, which we
call {\em lexicographic} order, as follows. Arrange the elements of each
part in the order inherited from $S$, then arrange the parts in the
order of their first elements
(this is the order introduced in \cite[\S8.3.2]{Martin91}). 

Let us assume that $\St{3}[F,F']$ comes equipped with a total order
on  $F \dotcup F'$. This induces a total order on the components of
any diagram $D$ in $\Smin{3}[F,F']$ as follows. 
These components are in natural correspondence with the parts of
$p(D) \in \Pa(F \dotcup F')$,
and these are ordered by the lexicographic order above as
derived from the order on $F \dotcup F'$. 
We will write $f_{lex}$ for the numbering of components by this total
order. 


Now consider a \CD\ $D \in \St{3}[F,F']$. 
We may  regard the corresponding  
ordered pair of graphs $(\gtree(F),\gtree(F'))$ as
 a single graph $\gtree(F \dotcup F')$ by identifying the roots. 
Suppose that $D$ has $m$ connected components, and let 
$f $ be a map counting these components
(we say component $d$ has `colour' $f(d)$). 
We may associate the pair $(D,f)$ with a colouring 
$\phi(D,f)$ of the edges of
 $\gtree(F \dotcup F')$ 
as follows.
If boundary loop $l$ belongs to component $d$ in $D$ then 
the edge associated to loop $l$ is coloured by the colour $f(d)$. 

\begin{figure}
\[
\includegraphics{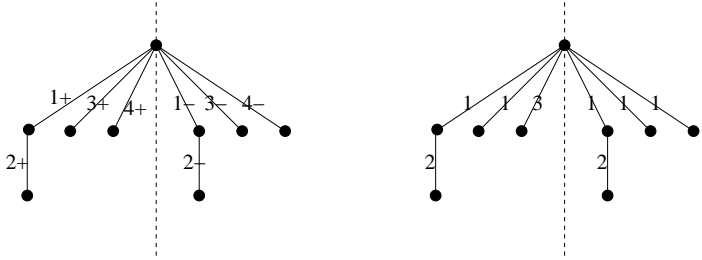}
\]
\caption{\label{GFF}(i) Edge/label correspondence; (ii) colouring.}
\end{figure}

\ex{Consider Figure~\ref{f:parti1}, which is of a diagram with three
  components, and take $f$ to enumerate these components in order of their
  leftmost points as drawn. The graph  $\gtree(F \dotcup F')$ is as
  shown on the left in Figure~\ref{GFF}. The colouring $\phi(D,f)$ is
  as shown on the right.}

Note that $D \simsh D'$ implies $\phi(D,f_{lex})= \phi(D',f_{lex})$. 
Thus $\phi$ can be considered as defined on $\she$-classes of diagrams. 
Indeed it is an injective map from  $\she$-classes into 
$\gtree(F \dotcup F')^{\{ 1, 2, ..., n \}}$,
the image of which then defines a partition, via the argument
following Definition~\ref{col graph}.
Thus we will have the desired characterisation of  $\St{3}_{\she}$
if we can describe the image of $\phi$. 


For any two edges $e,e'$ in a tree let $\chain(e,e')$ denote the chain
of edges connecting them in the tree (excluding $e$ and $e'$
themselves). 
\de{
An element of $\gtree(F \dotcup F')^C$ is said to be admissible iff
for every pair of same coloured edges $e,e'$ either there is another
edge in $\chain(e,e')$ also of the same colour, or every colour
appearing in $\chain(e,e')$ occurs an even number of times. 
}
For example, the colouring on the right in Figure~\ref{GFF} is
admissible. The only non-empty chain to check is 
between the two edges with the colour
2. Here the chain has colour sequence 1,1 (which is an even number of
each colour). Meanwhile replacing either 2 with a 3 would make an
inadmissible colouring, since then the two edges with colour 3 would
define a chain containing a single 1.

\pr{
The images under $\phi$ of coloured \CD s $D$ with $n$ components 
and $D \in \St{3}[F,F']$ 
are precisely the  admissible graph colourings in 
$\gtree(F \dotcup F')^{\{ 1, 2, ..., n \}}$.
}
\proof{We need to show (I) that every $\phi(D,f)$ is admissible;
and (II) that every admissible graph colouring is the
image of some $D$. 

(I):
Consider any  $\phi(D,f)$. 
We \RTS\ that whenever 
$f(e)=f(e')$ and 
no other edge in $\chain(e,e')$ has colour $f(e)$, then any given colour
occurs an even number of times.
But each component $d$ of $D$ partitions the remainder of the universe
in to two parts (with every other component entirely on one side or
the other). 
Thus, in passing from $e$ to $e'$, every time we pass 
through an edge of some colour we toggle the state of being inside or
outside of the corresponding component of $D$. 
Since every component lies entirely inside or outside every other in this
sense,  the fact that $e$ and $e'$ are in the same component implies
that any other given colour must appear an even number of times in the chain.  

(II) 
Consider any admissible graph colouring. 
Note that if there are two edges with the same colour then there are
two {\em adjacent} edges with the same colour. 
A \CD\ $D_1$ in which such adjacent edges are part of the same
component can be built by bridging these components in any 
\CD\ in which 
 each component of $D$ is a topological disk (with
boundary some component of $F \dotcup F'$).

If there are two edges with the same colour that are not adjacent
then there are two such separated only by adjacent same-colour pairs.
The corresponding loops are in different components in $D_1$. 
However these components
can be bridged in $D_1$ by a route which passes down each of
the bridges corresponding to the separating adjacent same-colour pairs. 

It is an exercise to show that this procedure can be
iterated.
}

\subsection{Examples}
\begin{figure}
\includegraphics{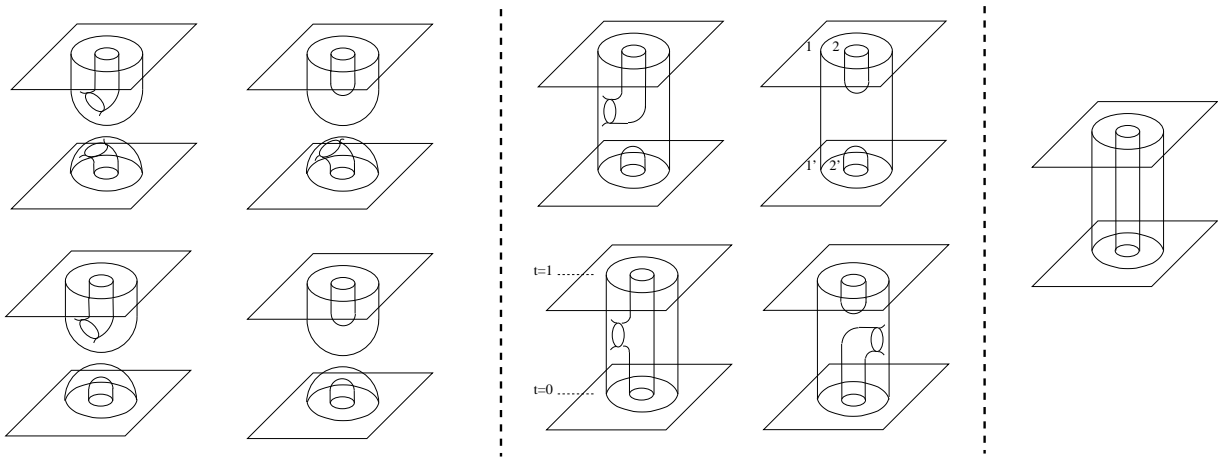}
\caption{\label{f:(())}  
Diagrams organised by increasing number of `propagating loops'
(i.e. connected components meeting both boundary hyperplanes).}
\end{figure}
The complete list of diagrams in $\St{}_{\she}[F,F]$ in case
$F=(())$ is given in Figure~\ref{f:(())}.
The best way to organise these is to note that the diagrams not only
form a basis for a $\K$-algebra, but also that this basis reveals a
sequence of ideals in the algebra. 
This follows from the fact that the product of two
diagrams can never be a diagram further to the right, as ordered in
the figure. (This is the analogue of the propagating line filtration
for the Temperley-Lieb algebra \cite{Martin91}.
From left to right in the figure we have first the diagrams with no
propagating loops; then those with one propagating loop; then those
with two.) 
This means that we do not need to compute the whole $9\times 9$
multiplication table to determine the structure of the algebra. 
It is enough to  compute within the sections of this filtration. 

The grouped diagrams are bases for the sections in a sequence of
double-sided ideals (i.e. for bimodules).
As {\em left}-modules 
(i.e. acted on by diagram multiplication from above)
these break up further -- as a direct sum of
isomorphic left-modules with bases given by the rows in each group
of diagrams. 
Let us restrict attention here to representation theory over the
complex field. Then over the ring $\C[\dq,\kp]$
these modules have an inner product defined on them via
duality (diagram inversion) and composition in the algebra.
The Gram matrices are 
\[
M(\emptyset)=\mat{cc} \dq^2&\dq \\ \dq&\dq\kp \tam ; \qquad
M(())=\mat{cc} \dq& 1 \\ 1& \kp \tam  ; \qquad 
M((()))=(1)
\]
where the argument is the {\em propagating} loop configuration. 
By the usual theory of Gram determinants \cite{MartinSaleur94a}
this shows that the algebra is generically semisimple, but
non-semisimple when 
\[
\dq (\dq\kp -1)=0 . 
\]
Note that this nicely generalises the 2D case (cf. \cite{Martin91}).
This raises many interesting questions (about the connection with
quantum groups \cite{Kassel95} for example), 
which will be treated elsewhere.


\section{Discussion}

\TL\ representation theory  controls the kind of observables 
and correlations occurring in certain physical models in
two-dimensions, as already noted. Via an appropriate limit it  
is also closely related to associated conformal algebras 
and their generalisations \cite{KooSaleur93}. 
At the same time, the
 category theoretic setting makes \TL\ representation theory  
{\em per se} relatively easy to analyze (see \cite{Martin91}). 
While the corresponding 
physical associations in three-dimensions
remain an intriguing open question for now, the matching categorical
structure (that we have introduced in this paper) does facilitate 
immediate progress in representation theory. 
We will compartmentalize the
construction (here) and the representation theory 
 (in a separate paper), however, as the representation theory 
constitutes an interesting 
(and rather long)
story in its own right. 

\bigskip
\noindent {\bf Acknowledgments} \qquad
We thank Anton Cox for helpful discussions, and Steffen Koenig
and the organisers of the First (Oxford) 
Diagram Algebras Conference, for giving MA the opportunity to
present these results there. 

\appendix   \[ \]
{\LARGE Appendix} 

\section{Rooted tree combinatorics} \label{A:tree}

We have seen in Propositions~\ref{pr:bcrt} and~\ref{pr:isoisot}  
that the object set for our category is given by the set
of rooted trees. This set thus takes the role played by the natural
numbers in the Temperley--Lieb category. Accordingly it will be 
useful to enumerate, order, and suitably partially order this set.


\medskip

\begin{figure}
\includegraphics{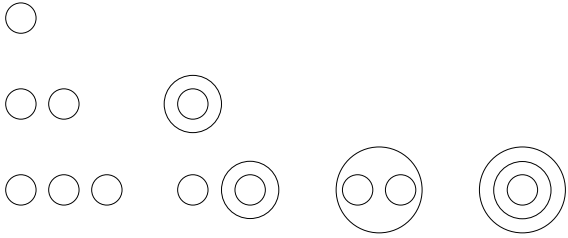}
\caption{\label{123} Loop configurations up to $n=3$.}
\end{figure}

The number $L_n$ of isotopy 
classes of loop configurations with $n$ loops 
(or equivalently of rooted trees)
is given by the generating functional 
\cite[\S3.17]{Wilf94} 
\begin{equation}
\sum_{n\geq 0} L_n x^n
=
\prod_{k\geq 1}(1-x^k)^{-L_{k-1}}
\] \[
=(1-x)^{-1}(1-x^2)^{-1}(1-x^3)^{-2}...
= (1+x+x^2+x^3+x^4...)(1+x^2+x^4+...)(1+2x^3+...)...
\] \[
=1+x+2x^2+4x^3+9x^4+...
\label{eq:gefu}
\end{equation}
In our heuristic bracket notation the first few are: 
$\;\; \emptyset, \;\; (), \;\; ()(), (()), $ 
$\;\; ()()(), ()(()), (()()), ((())), $
\[
\qquad
()()()(), ()()(()), ()(()()), ()((())), 
(()()()), (()(())), ((()())), (((()))), (())(())
\]
(we will give a more usable notation shortly); 
while the first few loop configurations are given in Figure~\ref{123}. 

\subsection{Ordering rooted trees} \label{B:tree}

Let $S$ be a set of symbols. 
The set of all  sequences in the set of symbols $S$ is 
denoted $\langle  S \rangle$. 
Consider the
the  two symbol `alphabet' 
$$
S_2=\{ \; ), \;  ( \; \} .
$$
Then for example, the sequence
$\; ))))(()(( \;\; \in \langle  S_2 \rangle$. 
A properly nested bracket sequence is any such bracket sequence in which
the running total of )'s never exceeds that of ('s.
The set $\bracket_n$ of nestings of $n$ bracket pairs is the set of
properly nested bracket sequences in which there are $n$ of each type
of bracket. 

\bigskip

A forest is a rooted tree with at least one vertex
(the point being that by removing the root we get a collection of
rooted trees --- each with root a child of the original root). 

It will be convenient to be able to totally order $\tree$, and hence
the set of forests. 
To this end we first introduce a larger set of trees, with a natural 
total order.

\de{
A rooted plane tree is a rooted tree with an ordering for the children
of each vertex. 
\\
Write $\ptree$ for the set of rooted plane trees and $\ptree_n$  
for the set of rooted plane trees with $n$ vertices.
}
The child ordering passes lexicographically to an ordering of all
vertices of a rooted plane tree 
(root first; then the first child of the root; then her first child 
(else the second child of the root);
and so on). 
\\
The {\em depth} of a vertex is the distance from the root.


The set $\ptree$ of rooted plane trees, 
and the  set $\ptree_n$ of rooted plane trees with $n$ vertices, 
may each be ordered by the lex order on the depth sequences
of their elements.


In order to illustrate this order it will be convenient to have an inline
representation of rooted plane trees. 
Let $R$ be any planar representation of $t \in \ptree$ in the upper
half-plane,
such that the root lies at (0,0), and the children of each vertex
appear in clockwise order around it.
The {\em traversal} of $R$ is the sequence of directed edges 
which must be traversed to  move clockwise round
the outside of the tree from root back to root. The map $\tau$ from $R$ to
sequences in the two symbol alphabet $\{ ), ( \}$ takes $R$ to the sequence 
recording ( for each outward directed edge and ) for each directed
edge back towards the root. 
It will be evident that $\tau(R)$ depends only on $t$. 

\pr{
The map $\tau$ defines a bijection  
$\tau:\ptree_n \rightarrow \bracket_n$. 
}
Using the inline representation of rooted plane trees 
thus provided by bracket sequences
(but augmented by an extra outer layer of brackets so that the root
may be explicitly given its depth 0 label), 
the abovementioned order on $\ptree$ begins:
\[
\begin{array}{llllllllll}
() & (()) & (()()) & ((())) & 
    \; (()()()) & (()(())) & ((())()) & ((()())) & (((())))
\\
0 & 01 & 011 & 012 & 0111 & 0112 & \underline{0121} & 0122 & 0123
\end{array}
\]


There is a map from rooted plane trees to rooted trees which simply
forgets all the child labels. 

\de{ \label{heavy}
Let $T_1,T_2$ be two rooted plane trees. If the sequence of depths 
(of vertices in the traversal -- including only the first occurence of
each, as exemplified above) 
first differ in a vertex of $T_1$ deeper than one of $T_2$ we say $T_1$ is
heavier than $T_2$ \cite{NakanoUno03}. 
}

Among rooted plane trees corresponding to the same rooted tree there
is one which is heavier than no other. 
The set of these {\em left-light} rooted plane trees thus provides a
unique such tree for each rooted tree. 

The set of rooted trees (and hence the set of forests) 
inherits an order from the subset order on the
subset of  {\em left-light} rooted plane trees. 

\subsection{Partial orders on the set of rooted trees}

\newcommand{\subtree}{\lhd}
\newcommand{\subtreeeq}{\unlhd}
\newcommand{\suptree}{\rhd}
\newcommand{\suptreeeq}{\unrhd}
\newcommand{\subfold}{\lhd_f}
\newcommand{\subfoldeq}{\unlhd_f}
\newcommand{\supfold}{\rhd_f}
\newcommand{\supfoldeq}{\unrhd_f}

Let $t_1,t_2$ be bracket sequences such that the concatenation $t_1 t_2$
is a matched bracket sequence (and hence a notation for a rooted plane tree).
Then if $t$ is another  matched bracket sequence then 
$t_1 t t_2$ is another (and hence gives another such tree). 
Note for example that the set of sequences of form $(t)$ has no
element
in common with the set of form $()t$. 

A rooted subtree of a rooted tree is one which can be obtained by (iterated)
removal of leaves.
We write $a \subtree b$ if $a$ is a rooted subtree of $b$.
An important {\em partial} order on the  set of rooted trees (and
hence the set of forests)  is the rooted-subtree order.

The {\em fold} operation on a rooted tree is of form
\[
(t_1(t_2)) \mapsto (t_1)t_2
\]
where $t_1$ and $t_2$ are any trees 
(matched bracket sequences in this notation),
and may be applied anywhere in a tree matching this pattern.
\\
Example: 
\eql(eq:fold)
((())) \mapsto ()() .
\eq

A {\em meld} operation is of form 
\[
(t_1)(t_2) \mapsto (t_1 t_2)
\]
Example: $(())(()) \mapsto (()())$.

\de{(Sub/fold order)
We write $a \subfold b$ if $a$ may be obtained by $b$ by any sequence
of fold and meld  operations and leaf removals.
}

The following figure is the beginning of the Hasse diagram for the 
sub/fold order:
\[
\xymatrix{
                 &                    & ()() \ar[r]\ar[dr]\ar[dddr] & ()()()
\\
\emptyset \ar[r] & () \ar[ur] \ar[dr] &                             & ()(())
\\
                 &                    & (()) \ar[ur]\ar[r]\ar[dr]   &
                 (()())
\\
                 &                    &                             & ((()))
}
\]
We are describing this order because it turns out to play a big role
in the representation theory of our categories (as we outline in
section~\ref{S:diagcat}). 

Note that this poset is not a lattice, in particular it does not have a 
meet: ((())) and ()(()) `meet' at (()) and ()().

\section{Diagram category representation theory}\label{S:diagcat}

Here we give  a very brief preview of the organisational scheme now
available to us for analysing the reductive representation theory of our
diagram categories
(i.e. the search for simple modules of the
diagram algebras contained therein). 

In this section  $C=(S_C,\hom_C(-,-),\circ)$ is a category, 
$\hom_C$ denotes the  class of all morphisms in $C$,
and $\One_F$ the identity hom in $\hom_C(F,F)$.
Here we use the `diagram' notation for homs, meaning that they compose in
the order 
$\hom_C(F',F) \times \hom_C(F,F'') \rightarrow \hom_C(F',F'')$.
We assume (merely for notational simplicity) that all 
our categories are small. 

For any poset $(T,\leq)$ and map $f: \hom_C \rightarrow T$ 
we say that $C$ is filtered by $f$ if for each composable
pair of homs $D,D'$ we have
$
f(D \circ D') \leq f(A)  \mbox{ for } A \in \{D,D' \} . 
$


\ex{\label{ex:pn}
For $D \in \St{d}$ (any $d$)
the propagating number $\#(D)$ is simply the number of
components of $D$ that contain boundary components in both boundaries.
We have
\[
\#(D \circ D') \leq \mbox{min}(\#(D ),\#( D'))
\]
so $\CC{d}$ is filtered by $\#$.
} 
In a $\K$-linear category with a given 
collection of bases we 
will adopt the convention that such a filter, if defined on the bases,
takes the lowest value on linear combination $X$ from the basis elements
with finite support in $X$.
Then $\FF(\CC{d})$ and $\CC{}_{\she}$ and $\CC{}_{\he}$
are also filtered by $\#$. 


\de{ \label{de:facto}
A morphism $D$ in a category $C=(S_C,\hom_C(-,-),\circ)$
{\em factors through} object $F \in S_C$
if $D = D' \circ D''$ with $D' \in \hom_C(-,F)$ and  $D'' \in \hom_C(F,-)$.
\\
For each partial order  $\preceq$ on object set $S_C$ then 
$\#^{\preceq}(D)$ is the
set of $\preceq$-lowest objects in $S_C$ that $D$ factors through.
}
\ex{
Consider the \CD s in Figure~\ref{f:tworealisations} as 
representatives of homs in $\CC{}_{\she}$.  
The left diagram 
(call it $[L]_{\she}$)
factors through objects in the isotopy class (()),
and also through objects with more loops, but not through any 
object in the class ()(), nor any object with fewer loops.
The right diagram factors through ()(). 
}


\de{ With the setup of Definition~\ref{de:facto},
we say category $C$ is filtered by  $\preceq$ if 
$F \in \#^{\preceq}(D \circ D')$ implies $F \preceq F'$ for all 
$F' \in \#^{\preceq}(D ) \cup \#^{\preceq}(D')$.
}

The point about such a filter, when it exists, is that it leads to a
filtration on ideals, and hence to 
an initial decomposition in representation theory.
This raises the question of how to construct such a filter,
which in our case has a rather neat answer. 


\newcommand{\propags}{\geq_p}

\de{For each category $C$ 
define a relation on $S_C$ by $F \propags  F'$ if the map
\eqal()
\hom_C(F',F) \times \hom_C(F,F') & \rightarrow & \hom_C(F',F')
\\
(A,B) & \mapsto &  A \circ B
\eqa
is surjective. That is to say,  $F \propags  F'$ if $\One_{F'}$ factors
through $F$. 
}

\pr{
The relation $\propags$ is reflexive and transitive, for any category,
but not in general antisymmetric.
}
If the relation is antisymmetric we call poset $(S_C,\propags)$ the 
{\em propagating order} on $C$.


\ex{
(i) Any skeleton of the categorical basis $\CC{}_{\she}(1,1)$ 
has a propagating order. 
(ii) This order is given by the 
sub/fold order on rooted trees (and hence on objects).
}
{\em Outline proof:}
(i) By \ref{ex:pn}  $ F \propags F'$ implies $|F| \geq |F'|$.
If  $ F \propags F'$ and  $|F| = |F'|$ then there is a factorisation
through $F$ of form $1_{F'} = D \circ D'$. But then $D' \circ D$ is 
an isomorphism, and hence so is $D$,  
and $F=F'$ by the skeleton property. 
\\
(ii) We need to show that the propagating relation is given by the sub/fold
order. 
It will be clear that the sub order is a subrelation
(i.e. that $ F \suptree F' $ implies $ F \propags F'$).
Now suppose that $F$ and $F'$ differ by a fold operation. 
Locally the propagating condition is satisfied by the following
composition, 
which shows that $\One_{F'}$ with $F' \sim ()()$ factors 
with $F \sim ((()))$ in the middle layer
(noting that  $\kp=1$ is a unit):
\[
\includegraphics{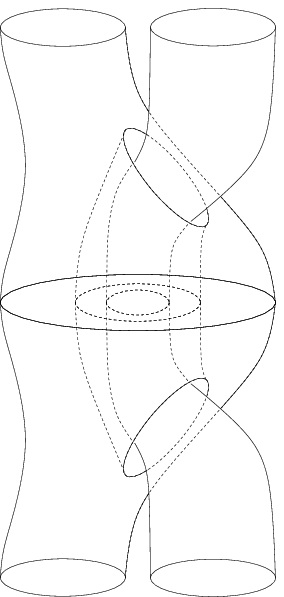}
\]
Now compare with equation~(\ref{eq:fold}).
A similar picture can be drawn for the meld operation. 
It follows from this that  
$ F \supfold F' $ implies $ F \propags F'$. 
(We will complete the argument elsewhere.)


A propagating order has very useful consequences in 
representation theory. Their specific development depends on whether the 
category has a $\Ring$-linear structure.
For the sake of simplicity here we assume it does not.
See \cite{Martin07} for the $\Ring$-linear case. 

\de{
Write $\hom_C^G(F,F')$ for the subset of $\hom_C(F,F')$ of homs that
factor through $G$. 
}

Note that $\hom_C(F,F')$ is an $M$-set for the monoid $M = \hom_C(F,F)$,
by the category composition. 

\pr{
If $G \propags G'$ then  $\hom_C^G(F,F') \supseteq \hom_C^{G'}(F,F')$
is an inclusion of $M$-sets.
\Qed}

\de{
If $C$ has a propagating order then  
$$
\hom_C^{=G}(F,F') = \hom_C^G(F,F') \setminus \cup_H \hom_C^H(F,F')  
$$
where the union is over all $H $ below $G$ in the order. 
}
Note that $\hom_C^{=G}(F,F')$ is empty unless $ F,F' \propags G$.

\ex{ Consider 
 $C=\C_{\she}(1,1)$, and $F\sim (())$. Then $\hom_C(F,F)$ is shown in
 Figure~\ref{f:(())}. The set $\hom_C^{=()}(F,F)$ is the middle group of four
 diagrams. The set $\hom_C^{=\emptyset}(F,F)$ is the leftmost group of
 four, and
\[
\hom_C^{()}(F,F) = \hom_C^{=()}(F,F) \cup \hom_C^{=\emptyset}(F,F)
\]  
}


The next question is whether this ideal structure can be further
refined. In particular what is the structure of $\hom_C^{=F}(F,F)$?
In our case we already have the $\#$ filtration to hand,
so we should next bring the two structures together. 

\de{
A \CD\ $D \in \St{d}[F,F']$ is {\em full} on $F$ 
(\resp\ $F'$) if $\#(D)=|F|$ (\resp\  $\#(D)=|F'|$). 
\\
$\St{d}_{F}[F,F']$ (\resp\  $\St{d}_{F'}[F,F']$)
is the subset full on $F$ (\resp\ $F'$). 
}
\pr{
Let $C$ be a skeleton of $\C_{\she}(1,1)$.
If $D \in \St{d}[F,F']$ full on $F$  
then $\#^{\subtreeeq_f}(D)= \{F \}$.
In particular if $\#(D) = |F| = |F'|$ then $F=F'$.
}
\newcommand{\opp}{\overline}
\proof{
Let $D=D' \circ D''$ be any factorisation of $D$, through $G$ say.
Then $D'$ is full on $F$, and $D' \circ \opp{D'}$ 
($\opp{D'}$ is the  upside-down version of $D'$) is congruent to
$1_F$,
so that $G \propags F$. 
Object $F$ itself is thus uniquely lowest among such $G$s.  
}

The corresponding observation for the TL category determines its
representation theory in large part. We will investigate the
present case in detail in a separate work.


\appendixCC

\bibliographystyle{amsplain}
\bibliography{new31,main,emma}

\end{document}